\documentclass[pdflatex,sn-mathphys-num,iicol]{sn-jnl}

\usepackage{caption}

\usepackage{graphicx}%
\usepackage{multirow}%
\usepackage{amsmath,amssymb,amsfonts}%
\usepackage{amsthm}%
\usepackage{mathrsfs}%
\usepackage[title]{appendix}%
\usepackage{xcolor}%
\usepackage{textcomp}%
\usepackage{manyfoot}%
\usepackage{booktabs}%
\usepackage{listings}%
\usepackage{amssymb}
\usepackage{amsmath}

\usepackage{todonotes}

\usepackage[shortlabels]{enumitem}

\usepackage{pifont}
\newcommand{\cmark}{\textcolor[rgb]{0.00,0.59,0.00}{\ding{51}}}%

\usepackage{graphicx,xcolor}
\definecolor{gainsboro}{rgb}{0.86,0.86,0.86}

\usepackage{array,multirow}
\usepackage{multicol}

\usepackage{lineno,hyperref}
\modulolinenumbers[5]
\hypersetup{
    colorlinks=true,
    linkcolor=blue,
    filecolor=magenta,
    urlcolor=cyan,
    pdftitle={Overleaf Example},
    pdfpagemode=FullScreen,
}

\usepackage{colortbl}
\definecolor{papayawhip}{rgb}{1.0, 0.94, 0.84}
\definecolor{midnightgreen}{rgb}{0.16863, 0.42353, 0.65098}
\usepackage[skins]{tcolorbox}
\tcbset{summary/.style={colback=papayawhip,colframe=midnightgreen, width=\linewidth,nobeforeafter}}

\usepackage{caption}
\usepackage{subcaption}

\definecolor{red_listing}{RGB}{233,30,99}
\usepackage{listings}
\lstdefinelanguage{KCONFIG}{
    otherkeywords={
        !, &&
    },
    morekeywords={config, bool, depends, on, !},
    sensitive=true,
    morestring=[b]" 
} %
\usepackage[noend,]{algpseudocode}
\usepackage{algorithm,algorithmicx}

\algnewcommand{\LineComment}[1]{{\color{blue}\Statex \(\triangleright\) {#1}}}

\usepackage{amsthm}
\usepackage{thmtools}

\declaretheoremstyle[
    headfont=\bfseries,
    notefont=\bfseries\itshape,
    notebraces={(}{)},
    bodyfont=\normalfont,
    headpunct={},
    postheadspace=\newline,
    postheadhook={\textcolor{MiAzul}{\rule[.6ex]{\linewidth}{0.4pt}}\\},
    spacebelow=\parsep,
    spaceabove=\parsep,
    mdframed={
        backgroundcolor=MiAmarillo,
        linecolor=MiAzul,
        innertopmargin=3pt,
        roundcorner=5pt,
        innerbottommargin=6pt,
        skipabove=\parsep,
        skipbelow=\parsep }
]{mydefistyle}

\newcommand{\tool}[1]{\textsc{#1}}
\newcommand{\code}[1]{\textsf{#1}}
\newcommand{\sat}{\textsf{sat}}
\newcommand{\unsat}[1]{\textsf{unsat}}
\newcommand{\ddnnf}{\mbox{d-DNNF}}

\theoremstyle{thmstyleone}%
\theoremstyle{thmstyletwo}%

\theoremstyle{thmstylethree}%

\begin{document}

\title[Backbone Algorithms for Configurable Systems]{A Comparative Analysis of Backbone Algorithms for Configurable Software Systems}


\author[1]{\fnm{Luis} \sur{Cambelo}}\email{lcambelo1@alumno.uned.es}
\equalcont{These authors contributed equally to this work.}

\author*[1]{\fnm{Ruben} \sur{Heradio}}\email{rheradio@issi.uned.es}
\equalcont{These authors contributed equally to this work.}

\author[2]{\fnm{Jose-Miguel} \sur{Horcas}}\email{horcas@uma.es}
\equalcont{These authors contributed equally to this work.}

\author[1]{\fnm{Dictino} \sur{Chaos}}\email{dchaos@dia.uned.es}
\equalcont{These authors contributed equally to this work.}

\author[1]{\fnm{David} \sur{Fernandez-Amoros}}\email{david@issi.uned.es}
\equalcont{These authors contributed equally to this work.}

\affil*[1]{\orgname{Universidad Nacional de Educacion a Distancia (UNED)}, \orgaddress{\street{Calle Juan del Rosal, 16}, \city{Madrid}, \postcode{28040}, \country{Spain}}}

\affil[2]{\orgname{ITIS Software, Universidad de Málaga}, \orgaddress{\street{Blvr. Louis Pasteur, 35}, \city{Málaga}, \postcode{29071}, \country{Spain}}}


\abstract{The backbone of a Boolean formula is the set of literals that must be true in every assignment that satisfies the formula. This concept is fundamental to key operations on variability models, including propagating user configuration decisions to identify implied feature selections, detecting dead features and dead code blocks, and preprocessing formulas to accelerate knowledge compilation into tractable representations such as binary decision diagrams. Despite its importance, previous empirical studies have evaluated backbone algorithms solely on SAT competition formulas (typically engineered to test the limits of SAT solvers), leading to inconsistent conclusions about their performance. This study provides the first comprehensive evaluation of formulas derived from real-world variability models, analyzing 21 configurations of 5 state-of-the-art algorithms on 2,371 formulas from configurable systems ranging from 100 variables and 179 clauses to 186,059 variables and 527,240 clauses. The results indicate that variability model formulas are structurally distinct, with higher clause density but greater clause simplicity. Our research provides clear algorithm selection guidelines: Algorithm 2/3 (iterative with solution filtering) is recommended for formulas with 1,000 or fewer variables, while Algorithm 5 (chunked core-based) with adaptive chunk size selection provides the best practical performance for larger formulas. Also, the results show that filtering heuristics have negligible or negative effects on performance for variability models. Finally, the study identifies a research gap: while Algorithm 5 with optimal chunk size can achieve runtime reductions exceeding 50\% compared to Algorithm 2/3 (the one that product line tools implement), the optimal chunk size varies unpredictably across formulas and cannot currently be estimated, opening directions for future research. These findings offer actionable recommendations for tool developers, practitioners, and researchers.}

\keywords{variability models, configurable systems, software product lines, backbone, SAT solvers, empirical evaluation}



\maketitle

\section{Introduction}\label{sec:introduction}


\textit{Variability Models} (VMs) represent the features of configurable systems and the constraints that determine their valid combinations. For their automated analysis, these models are translated into Boolean formulas and subsequently processed by logic engines, e.g., SAT solvers~\cite{Batory05,She10,Fernandez19,Feichtinger21}. Each variable $x$ in a formula has two literals associated: $x$ and $\overline{x}$. The \textit{backbone} of a formula is defined as the set of literals that are \code{true} in every variable assignment that satisfies the formula; in other words, the literals that are common to all the variable assignments that render the formula \code{true}.

Backbones provide essential insights into the configuration spaces of VMs. Specifically, positive backbone literals correspond to \textit{core features} required in every valid configuration, whereas negative backbone literals correspond to \textit{dead features} that cannot be selected.

Beyond feature classification, backbone computation serves three essential purposes detailed in Section~\ref{sec:motivation}: (i)~propagating configuration decisions to reveal implied feature selections~\cite{Krieter18}, (ii)~detecting dead code blocks that can never be activated~\cite{Tartler12,Tartler13}, and (iii)~accelerating knowledge compilation into tractable representations such as \emph{Binary Decision Diagrams} (BDDs) or \emph{deterministic Decomposable Negation Normal Forms} (d-DNNFs)~\cite{Lagniez14,Lagniez2017,Kuiter25}.

\subsection{The Problem: Contradictory Prior Findings}

Given the importance of backbone computation, the SAT community has conducted extensive research on developing efficient algorithms. The seminal work by Janota et al.~\cite{Janota2015_Algorithms} introduced chunk-based algorithms that test multiple literals simultaneously, implemented in the \tool{MiniBones} tool. Zhang et al.~\cite{Zhang2020_EDUCIBone} proposed the \tool{EDUCIBone} tool, which incorporates novel literal filtering heuristics. Most recently, Biere et al.~\cite{Biere2023_CadiBack} presented \tool{CadiBack}, leveraging the modern \tool{CaDiCaL} SAT solver with transparent incremental processing.

However, the experimental findings from these studies are strikingly contradictory. They disagree on which algorithm performs best, whether chunk size should be small or large, and whether literal filtering helps, harms, or has no effect on performance (see Section~\ref{sec:algorithms} for details). These contradictions leave practitioners and tool developers without clear guidance.

A critical observation is that all three studies evaluated their algorithms exclusively on formulas from SAT competitions, which are general-purpose benchmarks designed to stress-test SAT solvers across diverse problem domains. No comprehensive empirical evaluation has examined backbone algorithm performance on formulas derived from real VMs (see Section \ref{sec:q1}). This gap is significant because VMs exhibit structural characteristics that differ fundamentally from SAT competition benchmarks.

Furthermore, there is a notable disconnect between the SAT and \emph{Software Product Line} (SPL) communities. While the SAT community has developed sophisticated chunk-based algorithms (Algs.~\ref{alg:all-in} and \ref{alg:all-out}) with various optimization techniques, popular SPL tools such as \tool{FlamaPy}~\cite{Galindo2023_Flama} and \tool{FeatureIDE}~\cite{Thum2014-FeatureIDE} implement only the simpler iterative approaches (Algs.~\ref{alg:naive-iterative} and \ref{alg:advanced-iterative}). To the best of our knowledge, the advanced algorithms from the SAT community have not been adopted or even discussed in SPL research.

\subsection{Our Contribution}

This paper bridges the gap between the SAT and SPL communities by presenting a comprehensive empirical study of backbone algorithms on formulas derived from real VMs. Our contributions are:

\begin{enumerate}[leftmargin=*]
    \item \textbf{Unified algorithm presentation.} We provide the first unified presentation of all five backbone algorithms, using consistent notation and terminology. This includes the advanced chunk-based algorithms (Algs.~\ref{alg:all-in} and \ref{alg:all-out}) that, despite being available since 2015, have not been previously discussed in the SPL literature.

    \item \textbf{Large-scale empirical evaluation.} We evaluate these 5 algorithms across 21 configurations on 2,371 formulas from real configurable systems, totaling 1,287,453 executions. Our benchmark includes systems ranging from small embedded tools to the Linux Kernel with 186,059 variables and 527,240 clauses.

    \item \textbf{Structural characterization.} We demonstrate that VM formulas are structurally distinct from SAT competition benchmarks: approximately 98\% of clauses contain at most two literals, compared to only 39\% in SAT competition formulas.

    \item \textbf{Algorithm selection guidelines.} We identify a clear size-dependent performance pattern: Alg.~\ref{alg:advanced-iterative}/\ref{alg:simplified-iterative} (iterative with solution filtering) is fastest for formulas with 1,000 or fewer variables, while Alg.~\ref{alg:all-out} (chunked core-based) with adaptive chunk size selection provides the best practical performance for larger formulas.

    \item \textbf{Filtering heuristic evaluation.} We show that filtering heuristics either have no measurable effect (rotatable literals) or actively harm performance (COV, WHIT, 2LEN) on VM formulas.

    \item \textbf{Practical tool.} We developed \tool{IPASIRBones}, a C++ implementation using the IPASIR interface that enables fair comparison across SAT solvers and provides practitioners with an efficient backbone computation tool for VMs with $\leq$ 1,000 variables.

    \item \textbf{Actionable recommendations.} We provide concrete guidance for tool developers (implement size-based algorithm selection and disable filtering), practitioners (make algorithm choice based on model size), and researchers (develop better chunk-size estimations).
\end{enumerate}

\subsection{Article Structure}

The remainder of this paper is organized as follows. Section~\ref{sec:motivation} presents three scenarios that highlight the practical importance of backbone computation in configurable systems. Section~\ref{sec:algorithms} describes the existing algorithms for computing backbones, their implementation in available tools, and the contradictory findings from prior experimental evaluations. Section~\ref{sec:experimental-evaluation} presents our empirical evaluation, addressing eight research questions about algorithm performance, the influence of chunk sizes, and the effect of available heuristics. The section concludes with a discussion of key findings, practical recommendations, and threats to validity. Finally, Section \ref{sec:conclusions} summarizes the main conclusions of our work.

\section{Three Situations that Highlight the Importance of Computing Backbones}\label{sec:motivation}

This section presents three scenarios where identifying the backbone is essential for addressing complex problems in configurable systems.

\subsection{Propagating Configuration Decisions}

The configurable features of a system and their relationships are typically represented with a variability model. For example, \tool{BusyBox}\footnote{\url{https://busybox.net/}}, is a software tool that replaces many standard GNU/Linux utilities with a single small executable, allowing for a customized environment suitable for multiple embedded systems. To achieve size optimization, \tool{BusyBox} is highly modular, supporting the selection of 613 features at compile time. \tool{BusyBox}'s variability model is defined with the \tool{Kconfig}\footnote{\url{https://www.kernel.org/doc/html/next/kbuild/kconfig-language.html}} language, which was originally devised for the \tool{Linux Kernel}\footnote{\url{https://www.kernel.org/}}, but nowadays is also used in many other systems \cite{Fernandez25}, such as \tool{Buildroot}\footnote{\url{https://buildroot.org/}},
\tool{Coreboot}\footnote{\url{https://www.coreboot.org/}}, \tool{Freetz}\footnote{\url{https://freetz.github.io/}}, etc.

Fig. \ref{lst:busybox-kconfig} shows a snippet from the \tool{Kconfig} specification of \tool{BusyBox}, which includes several \code{configs} that represent six features (\code{STATIC}, \code{PIE}, $\ldots$, \code{FEATURE\_SHARED\_BUSYBOX}) along with their interdependencies. All \code{configs} are Boolean (as indicated by the \code{bool} keyword in Lines 2, 4, ..., and 15), meaning they can either be selected or deselected. Each \code{config} includes a textual description to prompt the user on her preferences; e.g., Line 2 describes the meaning of including \code{STATIC} in a configuration. Additionally, specific dependencies between features are established; for instance, according to the \code{depends} statement in Line 10, the feature \code{BUILD\_LIBBUSYBOX} can only be selected if none of the following features are selected: \code{FEATURE\_PREFER\_APPLETS}, \code{PIE}, or \code{STATIC}.

\begin{figure*}[htbp!]
\begin{lstlisting}[language = KCONFIG]
config STATIC
  bool "Build BusyBox as a static binary (no shared libs)"
config PIE
  bool "Build BusyBox as a position independent executable"
  depends on !STATIC
config FEATURE_PREFER_APPLETS
  bool "exec prefers applets"
config BUILD_LIBBUSYBOX
  bool "Build shared libbusybox"
  depends on !FEATURE_PREFER_APPLETS && !PIE && !STATIC
config FEATURE_INDIVIDUAL
  bool "Produce a binary for each applet, linked against libbusybox"
  depends on BUILD_LIBBUSYBOX
config FEATURE_SHARED_BUSYBOX
  bool "Produce additional busybox binary linked against libbusybox"
  depends on BUILD_LIBBUSYBOX
\end{lstlisting}
\caption{Snippet of the \tool{BusyBox} \tool{Kconfig} specification.}\label{lst:busybox-kconfig}
\end{figure*}

An essential functionality of configurable systems is to propagate user decisions \cite{Krieter18}. For instance, if the user chooses to include \code{STATIC} in the configuration, the following \code{configs} must necessarily be excluded to ensure compliance with the variability model's constraints: \code{PIE}, \code{BUILD\_LIBBUSYBOX}, \code{FEATURE\_INDIVIDUAL}, and \code{FEATURE\_SHARED\_BUSYBOX}.

Note that, even in this very simple example, getting the propagation is not evident because of the transitive relations: \code{STATIC} excludes \code{BUILD\_LIBBUSYBOX} (Line 10 in Fig. \ref{lst:busybox-kconfig}), and \code{FEATURE\_INDIVIDUAL} requires \code{BUILD\_LIBBUSYBOX} (Line 13), so \code{STATIC} also excludes \code{FEATURE\_INDIVIDUAL} indirectly. This is an excerpt of a relatively small system, but there are very complex systems where propagating user decisions is extremely challenging, such as \tool{Linux Kernel v.6.8.4}, with 18,267 \code{configs} defined in 162,787 lines of \tool{Kconfig} code \cite{Fernandez25}.

To tackle decision propagation and many other relevant problems, variability models are first translated into Boolean formulas \cite{Batory05,She10,Berger13,Nadi14,Kastner17,Oh19tech,Fernandez19,Feichtinger21,Yaman23,Yaman24}, which are then processed using logic engines. For instance, Fig. \ref{lst:busybox-kconfig} is translated into the following formula $\varphi$:

\begin{footnotesize}
\begin{align*}
\varphi & \equiv (\overline{\mathrm{STATIC}} \vee \overline{\mathrm{PIE}})\wedge \\ \nonumber
& (\overline{\mathrm{BUILD\_LIBBUSYBOX}} \vee \overline{\mathrm{FEATURE\_PREFER\_APPLETS}})\wedge \\ \nonumber
& (\overline{\mathrm{BUILD\_LIBBUSYBOX}} \vee \overline{\mathrm{PIE}})\wedge \\ \nonumber
& (\overline{\mathrm{BUILD\_LIBBUSYBOX}} \vee \overline{\mathrm{STATIC}})\wedge \\ \nonumber
& (\overline{\mathrm{FEATURE\_INDIVIDUAL}} \vee \mathrm{BUILD\_LIBBUSYBOX})\wedge \\ \nonumber
& (\overline{\mathrm{FEATURE\_SHARED\_BUSYBOX}} \vee \mathrm{BUILD\_LIBBUSYBOX}) \nonumber
\end{align*}
\end{footnotesize}

For example, the $1^\mathrm{st}$ clause, $\overline{\mathrm{STATIC}} \vee \overline{\mathrm{PIE}}$, encodes the dependency specified in Line 5 of \ref{lst:busybox-kconfig}:  \code{PIE depends on !STATIC}. This relationship is translated as PIE $\Rightarrow$ $\overline{\mathrm{STATIC}}$, which is rewritten as the disjunction $\overline{\mathrm{STATIC}} \vee \overline{\mathrm{PIE}}$.

Once $\varphi$ is obtained, the propagation of a Boolean decision $d$ is computed as the backbone of $\varphi \wedge d$. The positive and negative literals in the backbone, denoted as $l$ and $\overline{l}$, indicate the \code{configs} that must be included in and excluded from the configuration, respectively. For example, selecting \code{STATIC} is propagated as:

        \begin{footnotesize}
        \begin{align*}
        \mathrm{Ba}&\mathrm{ckbone}(\varphi \wedge \mathrm{STATIC}) = \{\\ \nonumber
            & \mathrm{STATIC}, \\ \nonumber
        & \overline{\mathrm{PIE}},  \overline{\mathrm{BUILD\_LIBBUSYBOX}}, & \\ \nonumber
        & \overline{\mathrm{FEATURE\_INDIVIDUAL}},  \overline{\mathrm{FEATURE\_SHARED\_BUSYBOX}} \\ \nonumber
        \}& \\ \nonumber
        \end{align*}
        \end{footnotesize}

Note that this method supports propagating any number of decisions. For example, the propagation of including \code{BUILD\_LIBBUSYBOX}  and excluding \code{FEATURE\_PREFER\_APPLETS} is computed as:

        \begin{footnotesize}
        \begin{align*}
        \mathrm{Ba}&\mathrm{ckbone}(\varphi \wedge \mathrm{BUILD\_LIBBUSYBOX} \wedge \\ \nonumber
        & \overline{\mathrm{FEATURE\_PREFER\_APPLETS}}) = \{ \\ \nonumber
        & \ \ \mathrm{BUILD\_LIBBUSYBOX}, \overline{\mathrm{FEATURE\_PREFER\_APPLETS}}, \\ \nonumber
        & \ \ \overline{\mathrm{STATIC}}, \overline{\mathrm{PIE}} \\ \nonumber
        \} & \\ \nonumber
        \end{align*}
        \end{footnotesize}

\subsection{Identifying Dead Features and Dead Code}

As systems evolve, their variability models must stay in synchrony with all the remaining software artifacts by adding, removing, or modifying features and constraints. This evolution sometimes leads to problems. For instance, new constraints may render certain features \textit{dead}, meaning they cannot be included in any configuration because that would violate some variability model's constraints. This problem is not rare, e.g., among the systems analyzed in \cite{fernandez23}, \tool{Linux Kernel} has 4,766 dead features, \tool{CoreBoot} has 3,699, \tool{EmbToolkit} has 507, \tool{freetz} has 383, etc.

Once a variability model is translated into a formula $\varphi$, its dead features can be easily identified as the negative literals in the backbone of $\varphi$.

A related and more challenging problem is the detection of \textit{dead code}. To explain it, we will use an example taken from the \tool{Linux Kernel} \cite{Tartler2013_PhDThesis}. Fig. \ref{lst:dead-code-kconfig} shows part of a \tool{Kconfig} file that defines a selectable \code{config} called \code{USB\_HID}, which depends on two other \code{configs} named \code{USB} and \code{INPUT} (Line 4).  Fig. \ref{lst:dead-code-makefile} shows a fragment of a \textit{dancing makefile} \cite{Germaschewski03}, which indicates
file \code{usbhid.o} must be built if the user selects \code{USB\_HID}\footnote{\code{USB\_HID}'s type is \code{tristate}, meaning the user may decide to compile the feature statically (\code{y}), as a module (\code{m}) that will be loaded dynamically on demand, or not at all (\code{n}). Concerning Fig. \ref{lst:dead-code-makefile}, \code{USB\_HID}
 is selected when the user chooses \code{y} or \code{m}.}$^,$\footnote{In \code{makefiles} and \code{C} code, the flags referring to \code{configs} include a \code{CONFIG} prefix to distinguish them from other flags.}. \code{usbhid.o} is obtained by compiling \code{usbhid.c}, which is partially shown in Fig. \ref{lst:dead-code-c}. The preprocessor code decides what blocks (\code{B1}, \code{B2} or \code{B3}) must be compiled depending on the \code{configs} the user selects. The interesting aspect of this example is that, even though there are no dead features in the variability model, its combination with the part that generates the configuration does include dead code. Specifically, Block \code{B3} will never be generated for any configuration because to be activated in Fig. \ref{lst:dead-code-c}, \code{USB\_HID} should be selected and \code{INPUT} deselected (Lines 2 and 6). However, this scenario is impossible since, according to Fig. \ref{lst:dead-code-kconfig}, selecting \code{USB\_HID} requires \code{INPUT} to be selected as well (Line 4).


\begin{figure}[htbp!]
\begin{lstlisting}[language = KCONFIG]
config USB_HID
  tristate "USB HID transport layer"
  default y
  depends on USB && INPUT
  select HID
\end{lstlisting}
\caption{Snippet from the \tool{Kconfig} file \code{/drivers/hid/usbhid/Kconfig}.}\label{lst:dead-code-kconfig}
\end{figure}

\begin{figure}[htbp!]
\begin{lstlisting}[language = Make]
(CONFIG_USB_HID) += usbhid.o
\end{lstlisting}
\caption{Snippet from the \tool{Makefile} file \code{/drivers/hid/usbhid/Makefile}.}\label{lst:dead-code-makefile}
\end{figure}

\begin{figure}[htbp!]
\begin{lstlisting}[language = C]
#ifdef CONFIG_USB_HID
   // Block B1
#  ifdef CONFIG_INPUT
     // Block B2
#  else
     // Block B3
#  endif
#endif
\end{lstlisting}
\caption{Snippet from the \tool{C} file \code{/drivers/hid/usbhid/usbhid.c}.}\label{lst:dead-code-c}
\end{figure}

In this example, the dead code would be detected by translating Figs. \ref{lst:dead-code-kconfig} and \ref{lst:dead-code-c} to the formulas $\varphi_\mathrm{Kconfig}$ and $\varphi_\mathrm{c}$, respectively, and then computing the backbone of $\varphi_\mathrm{Kconfig} \wedge \varphi_\mathrm{c}$, which is $\{\overline{\mathrm{B3}}\}$, meaning that Block \code{B3} is always deactivated.

\begin{scriptsize}
\begin{align*}
\varphi_\mathrm{Kconfig} \equiv & (\overline{\mathrm{USB\_HID}} \vee \mathrm{USB})\ \wedge \\ \nonumber
& (\overline{\mathrm{USB\_HID}} \vee \mathrm{INPUT}) \\ \nonumber
\varphi_\mathrm{c} \equiv & (\overline{\mathrm{B1}} \vee \mathrm{USB\_HID})\ \wedge \\ \nonumber
& (\overline{\mathrm{B2}} \vee \mathrm{USB\_HID})\ \wedge \\ \nonumber
& (\overline{\mathrm{B2}} \vee \mathrm{INPUT})\ \wedge \\ \nonumber
& (\overline{\mathrm{B3}} \vee \mathrm{USB\_HID})\ \wedge \\ \nonumber
& (\overline{\mathrm{B3}} \vee \overline{\mathrm{INPUT}}) \\ \nonumber
\end{align*}
\end{scriptsize}

See \cite{Sincero10under,Dietrich12,Tartler12,Tartler13} for a detailed discussion on $\varphi_\mathrm{c}$ translation.

\subsection{Preprocessing Boolean Formulas for Knowledge Compilation}\label{sec:preprocessing}

The input of most Boolean logic engines is a formula in \textit{Conjunctive Normal Form} (CNF), which is a conjunction of one or more \textit{clauses}. Each clause is a disjunction of \textit{literals}, where a literal can be either a variable or a negated variable\footnote{All Boolean formulas presented so far are in CNF.}. For instance, the first column in Table \ref{tab:cnf-example} shows a CNF comprising six clauses, denoted as $\omega_i$.

For convenience, a CNF is sometimes represented as a set of clauses (as shown in the last column of Table \ref{tab:cnf-example}), where each clause is itself a set of literals. 


\begin{table}[!htpb]
\caption{Example of a CNF and its equivalent set representation.}
\label{tab:cnf-example}
\begin{center}
\begin{scriptsize}
\begin{tabular}{|c|c||c|}
\hline
\textbf{Boolean Notation} & \textit{Clauses} & \textbf{Set Notation}  \\ \hline \hline
$(\overline{a})\ \wedge$ & $\omega_1$ & $\big{\{}\ \ \{\overline{a}\},$ \\
$(\overline{a} \vee b)\ \wedge $ & $\omega_2$ & $\{\overline{a}, b\},$ \\
$(a \vee c)\ \wedge $ & $\omega_3$ & $\{a, c\},$  \\
$(\overline{c} \vee d) \wedge $ & $\omega_4$ & $\{\overline{c}, d\},$   \\
$(\overline{c} \vee e \vee f)\ \wedge$ & $\omega_5$ & $ \{\overline{c}, e, f\},$ \\
$(f \vee \overline{g})$ & $\omega_6$ & $\{f, \overline{g}\}\ \ \big{\}}$ \\ \hline
\end{tabular}
\end{scriptsize}
\end{center}
\end{table}

This set notation will be very useful in Section~\ref{sec:algorithms} for describing algorithms that compute backbones, and it will be used interchangeably with the logical notation. In particular, set union helps represent the addition of clauses and literals to $\varphi$, since $\varphi \cup S$ is equivalent to $\varphi \wedge S$. For example:

\begin{itemize}[leftmargin = *]
    \item $\varphi \cup \{l_1, l_2, \ldots, l_n\} \equiv \varphi \wedge l_1 \wedge l_2 \wedge \ldots \wedge l_n$
    \item $\varphi \cup \overline{\{l_1, l_2, \ldots, l_n\}} \equiv \varphi \wedge (\overline{l_1 \wedge l_2 \wedge \ldots \wedge l_n}) \equiv \varphi \wedge (\overline{l_1} \vee \overline{l_2} \vee \ldots \vee \overline{l_n})$
\end{itemize}

Since the number of configurations grows exponentially with the number of features a variability model has, many relevant analyses (e.g., finding configurations that optimize some performance indicators \cite{Oh23}, testing configurable software \cite{Mordahl19}, estimating each feature's influence on configuration performance \cite{Sincero10},  etc.) cannot be made on the whole population of configurations. Instead, they need to rely on a \textit{representative} random sample. A crucial requisite for representativeness is \textit{uniformity} \cite{Plazar19,Heradio22}, meaning that all configurations must have the same probability of being included in a random sample. All known sampling methods that guarantee uniformity require \textit{counting} the formula's number of solutions \cite{knuth09,Heradio22}.  In summary, reasoning about variability models often requires counting the solutions of their equivalent formulas.

Counting solutions of a CNF is known as the \#SAT problem, and it is computationally very expensive (it is \#P-complete, which is suspected to be even harder than NP-complete; see Chapter 25 in \cite{handbook}). Nevertheless, there are some \textit{knowledge compilation forms}, such as BDDs or \ddnnf{}s, which support counting in polynomial time \cite{Darwiche02}.

Unfortunately, the size of these normal forms is very sensitive to the variable and constraint ordering used in their synthesis, and finding an optimal ordering is itself NP-complete \cite{Bryant86}. Therefore, orderings are computed with many different heuristics (see Chapter 5 of \cite{Wegener87}). A complementary approach to ordering heuristics is preprocessing the input CNF to facilitate synthesizing the knowledge compilation forms. One of these preprocessing techniques uses the backbone of a CNF to simplify it as follows:

\begin{enumerate}[leftmargin=*]
    \item Add all literals in the backbone as unit clauses to the CNF.
    \item If a backbone literal appears in a clause, remove that clause from the CNF (i.e., as the literal is always \code{true}, the clause must be \code{true} since it is a disjunction).
    \item If a backbone literal appears in a clause with the opposite polarity (for example, if a backbone literal $l$ appears in the clause as $\overline{l}$, or a backbone literal $\overline{l}$ appears as $l$), it should be removed from the clause. This is justified because such a literal is always \code{false}.
\end{enumerate}

For example, the backbone of the CNF in Table \ref{tab:cnf-example} is $\{\overline{a}, c, d\}$. Accordingly, the formula simplifies to:
\begin{scriptsize}
\begin{align*}
\varphi \equiv &  \ \overline{a}\ \wedge \\ \nonumber
& c\ \wedge \\ \nonumber
& d\ \wedge \\ \nonumber
& (e \vee f)\ \wedge \\ \nonumber
& (f \vee \overline{g}) \\ \nonumber
\end{align*}
\end{scriptsize}

This preprocessing technique combined with ordering heuristics has experimentally proven beneficial for synthesizing BDDs and \ddnnf{}s \cite{Lagniez14,Lagniez2017,Dubslaff24,Kuiter25,Hess25}. An exceptional case happens with the \tool{Freetz}'s CNF provided in \cite{fernandez23}, with 67,546 variables and 240,767 clauses. Using the state-of-the-art compiler \tool{d4}\footnote{\url{https://github.com/crillab/d4}}\cite{Lagniez17}, we tried to obtain its \ddnnf{} on an HP Envy laptop equipped with an Intel Core i9-11900H Processor running at 2.5 GHz. After 8,889.08 seconds (nearly 2.5 hours), \tool{d4} automatically halted without producing the \ddnnf{}. Instead, by preprocessing the CNF with the backbone, we managed to eliminate 24,931 clauses and shorten other 14,024 clauses. With this preprocessed CNF as input, \tool{d4} successfully completed the \ddnnf{} synthesis in just 62.93 seconds.

\section{Existing Approaches for Computing Backbones and Their Current Experimental Evaluations}\label{sec:algorithms}

\subsection{A SAT Solver: the Underlying Engine}\label{sec:sat-solver}

This paper focuses on procedures that compute backbones by repeatedly invoking SAT solvers. Although these methods are computationally expensive (indeed, they are \textit{Co-NP complete} \cite{Janota2015_Algorithms}), they are the only ones that can tackle the largest variability models reported in the literature (e.g., the models included in \cite{fernandez23,Sundermann24benchmark}). A promising alternative involves encoding variability models into  \textit{knowledge compilation forms}, such as BDDs and \ddnnf{}s. When these forms are available, the backbone can be computed in polynomial time relative to the number of variables in the formula \cite{bdd-perezmorago, Sundermann24}. However, synthesizing these forms remains an open research problem as there is no way to obtain them for very large systems yet \cite{fernandez23,Tobias24,Dubslaff24}. This issue is somewhat circular: if we had the knowledge compilation form of a variability model, we could compute its backbone efficiently; however, if we knew the backbone in advance, we could simplify the formula as shown in Section \ref{sec:preprocessing}, thus facilitating the synthesis of the knowledge compilation form. 

Backbone computation is not the only problem that can be addressed by repeatedly calling a SAT solver; this approach is also applied in areas such as model checking \cite{Bradley11} and the MaxSAT problem \cite{Martins15}, among others. \textit{Traditional SAT solvers} poorly handle these situations because they evaluate the formula once and terminate, requiring each new call to start from scratch. In contrast, \textit{incremental SAT solvers} retain valuable information, such as learned clauses and variable activities, between repeated evaluations of slight variations of the same base formula. Furthermore, the \textit{Incremental Library Track of the SAT competition} introduced the standard interface for incremental SAT solvers IPASIR\footnote{\url{https://github.com/biotomas/ipasir}}$^,$\footnote{IPASIR stands for \textit{\underline{R}eentrant \underline{I}ncremental \underline{S}AT solver \underline{API}} when read backwards, but when pronounced forwards, it sounds like ``IPA, Sir?'', a humorous reference to asking someone if they want an IPA (India Pale Ale) beer.} \cite{Balyo15}, which is supported by several prominent solvers, including CaDiCaL\footnote{\url{https://github.com/arminbiere/cadical}}, MiniSat 2.2\footnote{\url{https://github.com/niklasso/minisat}}, Glucose\footnote{\url{https://github.com/audemard/glucose}}, and CryptoMiniSat\footnote{\url{https://github.com/msoos/cryptominisat}}.

\begin{figure}[htbp!]
    \centering
    \includegraphics[width=1\linewidth]{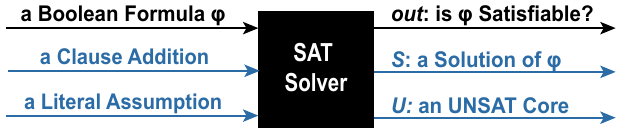}
     \caption{Inputs and outputs of incremental SAT solvers that implement IPASIR.}
     \label{fig:sat-solver-inputs-outputs}
\end{figure}

Fig. \ref{fig:sat-solver-inputs-outputs} shows the inputs and outputs of IPASIR solvers. The black arrows represent the most basic input and output: a solver takes a formula $\varphi$ in CNF and determines if it is \sat{} or \unsat{}. In addition, IPASIR offers the following extra input and outputs, depicted with blue arrows in Fig. \ref{fig:sat-solver-inputs-outputs}:

\begin{itemize}[leftmargin=*]
    \item Input:
    \begin{itemize}[leftmargin=6pt]
        \item The formula can be constrained by \emph{adding} new clauses.
        \item A literal $l$ can be \textit{assumed}, meaning the solver will consider $\varphi \wedge l$ in the next SAT search. Note that multiple literal assumptions can be chained before the search is performed, i.e., $\varphi \wedge l_1, \wedge l_2 \ldots \wedge l_n$. In contrast to clause addition, which is permanent, literal assumption is cleared after the search is done.
    \end{itemize}
    \item Outputs:
    \begin{itemize}[leftmargin=6pt]
        \item If $\varphi$ is \sat{}, the solver provides a \textit{solution} $S$, i.e., a complete assignment of variables that makes $\varphi$ \code{true} (\textit{complete} in the sense that \textit{all} the variables in $\varphi$ are assigned either \code{true} or \code{false}). For instance, the formula in Table \ref{tab:cnf-example} is \sat{}, and one of its solutions is $S~=~\{a$=\code{false}$,b$=\code{false}$,c$=\code{true}$,d$=\code{true}$,e$=\code{true}$,$ $f$=\code{false}$,g$=\code{false}\}, which we abbreviate as $S = \{\overline{a}, \overline{b}, c, d, e, \overline{f}, \overline{g}\}$.
        \item  If $\varphi$ is \unsat{}, the solver generates a proof of unsatisfiability known as an \textit{UNSAT core} and denoted as $U$. A core is a set of assumptions under which the formula is \unsat{}. While most modern solvers can identify relatively small cores, they cannot guarantee $U$ is \textit{minimal}.  This means that $U$ might still include assumptions not responsible for the conflict, i.e., assumptions whose removal would not make the core become \sat{}.

        For example, let us make three assumptions over the formula in Table \ref{tab:cnf-example}: $a$, $b$ and $\overline{c}$. As a result, $\varphi \wedge a \wedge b \wedge \overline{c}$ becomes \unsat{}. The solver could provide the core $U = \{a, b\}$, which is not minimal because, after removing $b$, $U = \{a\}$ is still \unsat{}; or perhaps, the solver could produce this other core $U = \{\overline{c}\}$, which is minimal. $U$ does not include all clauses that cause the formula unsatisfiability, but just a small bunch of clauses sufficient to render the formula \unsat{}; consequently, removing $U$ does not ensure the formula becomes \sat{}. For instance, even if $\overline{c}$ is eliminated, $\varphi \wedge a \wedge b$ remains \unsat{}.  
    \end{itemize}

\end{itemize}

\begin{figure*}[htbp!]
    \centering
    \includegraphics[width=0.8\linewidth]{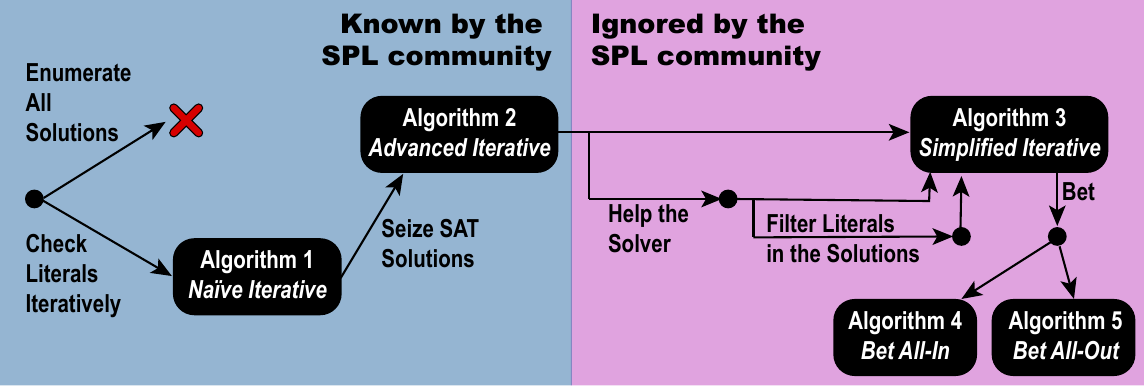}
     \caption{A map of backbone algorithms.}
     \label{fig:algorithms-map}
\end{figure*}

 When all the clauses in a CNF contain at most two literals, the SAT problem is called \textit{2-SAT} and can be solved in polynomial time (see Section 1.19 in \cite{handbook}). However, when there are clauses with more than two literals, the problem becomes NP. In Section \ref{sec:experimental-evaluation}, we will see that, although the Boolean translation of most variability models found in the literature consists primarily of clauses with two literals, they usually include a minority of clauses with more than two literals. Accordingly, the main strategy to improve the scalability of backbone computation is to reduce the number of SAT solver calls as much as possible. The following sections describe increasingly sophisticated approaches to accomplish this by seizing the inputs and outputs in Fig. \ref{fig:sat-solver-inputs-outputs}.

\subsection{Methods Used by the SPL Community for Computing Backbones}\label{sec:algorithms-spl}

Fig. \ref{fig:algorithms-map} shows the most relevant algorithms for computing backbones and their relationships. This section focuses on the blue area that covers the methods used by the SPL community. Section \ref{sec:algorithms-sat} discusses the more sophisticated approaches from the SAT community represented in the pink area.

\subsubsection{Enumerate All Solutions}

In 2007, Benavides \cite{Benavides07} pointed out that detecting core and dead features was one of the most relevant operations in the automated analysis of variability models. For computing them, he proposed the most straightforward approach: enumerating all configurations that comply with the variability model and then inspecting what features appear in all (core) and none (dead) of the configurations.

Once a variability model is translated to a CNF $\varphi$, whenever $\varphi$ is SAT, a solver can be used to enumerate all its solutions as follows:
\begin{enumerate}[leftmargin=*]
    \item A $1^\mathrm{st}$ solver call $\mathrm{SAT}(\varphi)$ provides a solution $S_1=\{l_1, l_2, \ldots, l_n\}$.
    \item A $2^\mathrm{nd}$ call asks for another solution different from the $1^\mathrm{st}$ one. This is done by adding a \textit{blocking clause} $\overline{S_1}$ to $\varphi$ (see Section 26.2.3 in \cite{handbook}):
    \begin{eqnarray*}
    \mathrm{SAT}\big{(}\varphi \cup \overline{S_1}\big{)} & \equiv \\
    \mathrm{SAT}\big{(}\varphi \wedge (\overline{l_1 \wedge l_2 \wedge \ldots \wedge l_n})\big{)} & \equiv \\
    \mathrm{SAT}\big{(}\varphi \wedge (\overline{l_1} \vee \overline{l_2} \vee \ldots \vee \overline{l_n})\big{)} & \equiv\\
    \mathrm{SAT}\big{(}\varphi \cup \{\vee_{l \in S_1}\overline{l}\}\big{)} &
    \end{eqnarray*}
    \item The process progresses until $m+1$ solver calls are made, all of them \sat{} except for the last one $\varphi \wedge \overline{S_1} \wedge \overline{S_2} \wedge \ldots \wedge \overline{S_m}$, which becomes \unsat{}. As a result, we have completed the enumeration of all the $m$ solutions of $\varphi$.
\end{enumerate}

Unfortunately, this method does not scale except for the tiniest variability models as it involves calling the solver as many times as solutions the formula has plus 1, and the number of solutions grows exponentially with the number of variables of $\varphi$. For example, the \code{Automotive02} CNF published in \cite{Krieter18} has 17,365 variables, giving a total of $5.260\cdot10^{1,441}$ solutions \cite{Heradio22}.

\subsubsection{Check Literals Instead of Solutions}

A formula $\varphi$ with $n$ variables has at most $2n$ literals (each variable $x$ has two associated literals: $x$ and $\overline{x}$) and $2^n$ solutions (each variable can be assigned to two values: \code{false} or \code{true}), as each variable can be set to 2 values: \code{true} or \code{false}. \tool{FlamaPy} \cite{Galindo2023_Flama} implements Algorithm \ref{alg:naive-iterative}, which takes advantage of the fact that $2n$ is much smaller than $2^n$. Instead of examining all solutions one by one, it checks whether each literal $l$ is part of the backbone $B$ as follows: $l$ is in $B$ if and only if $l$ appears in every solution of $\varphi$. Said the other way around, no solution includes $\overline{l}$. This can be verified by using a solver to demonstrate that $\varphi \cup \{\overline{l}\}$ is \unsat{}. When Algorithm \ref{alg:naive-iterative} finishes,  the affirmed literals $l$ in $B$ correspond to core features, while the negated ones $\overline{l}$ represent dead features.

\begin{algorithm}[htbp!]
\caption{\tool{Na\"ive Iterative}}
\begin{footnotesize}
\begin{flushleft}
\hspace*{\algorithmicindent} \textbf{Input}: A satisfiable formula $\varphi$ \\
\hspace*{\algorithmicindent} \textbf{Output}: The backbone $B$ of $\varphi$
\end{flushleft}
\begin{algorithmic}[1]
\State{\color{blue} $\triangleright$ $B$ will include the literals confirmed to be in the backbone}
\State $B \gets \emptyset$
\State{\color{blue} $\triangleright$ $C$ contains the literals that are candidates to be}
\State{\color{blue} in the backbone (they have not been tested yet)}
\State $C \gets$ literals$\big{(}\varphi\big{)}$
\While{$C \not= \emptyset$}
    \State $l \gets$ pick a literal from $C$
    \State $\mathrm{out} \gets \mathrm{SAT}\big{(}\varphi \cup \{\overline{l}\}\big{)}$ {\color{blue}\Comment{Is $l$ in the backbone?}}
    \If {$\mathrm{out}$ = \unsat{}} {\color{blue}\Comment{Yes}}
        \State $B \gets B \cup \{l\}$
    \EndIf
    \State $C \gets C \setminus \{l\}$
\EndWhile
\State  \Return{$B$}
\end{algorithmic}
\end{footnotesize}
\label{alg:naive-iterative}
\end{algorithm}

\subsubsection{Don't Ask What You Already Know}

\tool{FeatureIDE} \cite{Thum2014-FeatureIDE} implements the improved Algorithm \ref{alg:advanced-iterative}, which prevents the solver from being queried on literals whose backbone membership could be discarded in prior steps, thanks to the solutions generated by the solver.

Remember that Algorithm \ref{alg:naive-iterative} determines if a literal $l$ is part of $\varphi$'s backbone by testing the satisfiability of $\varphi \cup \{\overline{l}\}$ (which is the set-notation equivalent to $\varphi \wedge \overline{l}$). If this combination is \unsat{}, $l$ is in the backbone. Otherwise,  $l$ is not. However, when $\varphi \cup \{\overline{l}\}$ is \sat{}, the solver still provides valuable information: a solution $S$ that helps avoid future checks, as every literal $m$ in $S$ confirms that $\overline{m}$ is not in the backbone. Note that here we are using the same reasoning as Algorithm \ref{alg:naive-iterative}, but backward: $l$ is in the backbone if and only if $\varphi \wedge \overline{l}$ is \unsat{}; reversing the sentence, $\varphi \wedge \overline{l}$ is \sat{} if and only if $l$ is not in the backbone.

Algorithm \ref{alg:advanced-iterative} uses this extra information to:

\begin{itemize}[leftmargin=*]
    \item Reduce by half the initial list of candidates for the backbone with just a single solver call (Lines 2-5). This reduction occurs because each variable $x$ has associated two literals, $x$ and $\overline{x}$. Any solution includes only one of these literals, either $x$ or $\overline{x}$, but not both. Therefore, each solution contains half of the possible literals.
    \item Seize every solver call that yields \sat{} to increasingly filter out the candidate set (Lines 12-14).
\end{itemize}

\begin{algorithm}[htbp!]
\caption{\tool{Advanced Iterative}}
\begin{footnotesize}
\begin{flushleft}
\hspace*{\algorithmicindent} \textbf{Input}: A satisfiable formula $\varphi$ \\
\hspace*{\algorithmicindent} \textbf{Output}: The backbone $B$ of $\varphi$
\end{flushleft}
\begin{algorithmic}[1]
\State $B \gets \emptyset$
\State $C \gets$ literals$\big{(}\varphi\big{)}$
\State $(\mathrm{out}, S) \gets \mathrm{SAT}\big{(}\varphi\big{)}$
\State $\overline{S}\gets \{\overline{l}\ |\ l \in S\}$
\State $C \gets C \setminus \overline{S}$  {\color{blue}\Comment{$\overline{S}\not\subseteq B$, so avoid checking its literals}}
\While{$C \not= \emptyset$}
    \State $l \gets$ pick a literal from $C$
    \State $(\mathrm{out}, S) \gets \mathrm{SAT}\big{(}\varphi \cup \{\overline{l}\}\big{)}$   {\color{blue}\Comment{Is $l$ in the backbone?}}
    \If {$\mathrm{out}$ = \unsat{}}  {\color{blue}\Comment{Yes}}
        \State $B \gets B \cup \{l\}$
        \State $C \gets C \setminus \{l\}$
    \Else  {\color{blue}\Comment{No}}
        \State $\overline{S}\gets \{\overline{l}\ |\ l \in S\}$
        \State $C \gets C \setminus \overline{S}$  {\color{blue}\Comment{$\overline{S}\not\subseteq B$, so avoid checking its literals}}
    \EndIf
\EndWhile
\State  \Return{$B$}
\end{algorithmic}
\end{footnotesize}
\label{alg:advanced-iterative}
\end{algorithm}

\subsection{Advances Developed by the SAT Community}\label{sec:algorithms-sat}

\subsubsection{Helping the Solver and Filtering Literals in Solutions}\label{sec:filtering}

All literals in the backbone are present in every solution of the formula, so (i) the backbone is a subset of any solution ($B \subseteq S$), and (ii) the backbone is the intersection of all solutions ($B = {\bigcap}_{\forall S}S$). Using these two facts, Algorithm \ref{alg:simplified-iterative} rewrites Algorithm \ref{alg:advanced-iterative} in a more concise way, commonly used in the SAT literature \cite{Janota2015_Algorithms,Zhang2017a_DUCIBone,Biere2023_CadiBack}. In particular, Lines 2-5 in Algorithm \ref{alg:advanced-iterative} are rewritten with Line 2 in  Algorithm \ref{alg:simplified-iterative}, and Lines 13-14 with Line 14.

Fig. \ref{fig:set-equivalence} proves the most intricate equivalence between both algorithms: $C \setminus \overline{S} = C \cap S$. On the left, highlighted in yellow, the set of literals is divided into two complementary subsets:

\begin{figure*}[htbp!]
    \centering
    \includegraphics[width=0.8\linewidth]{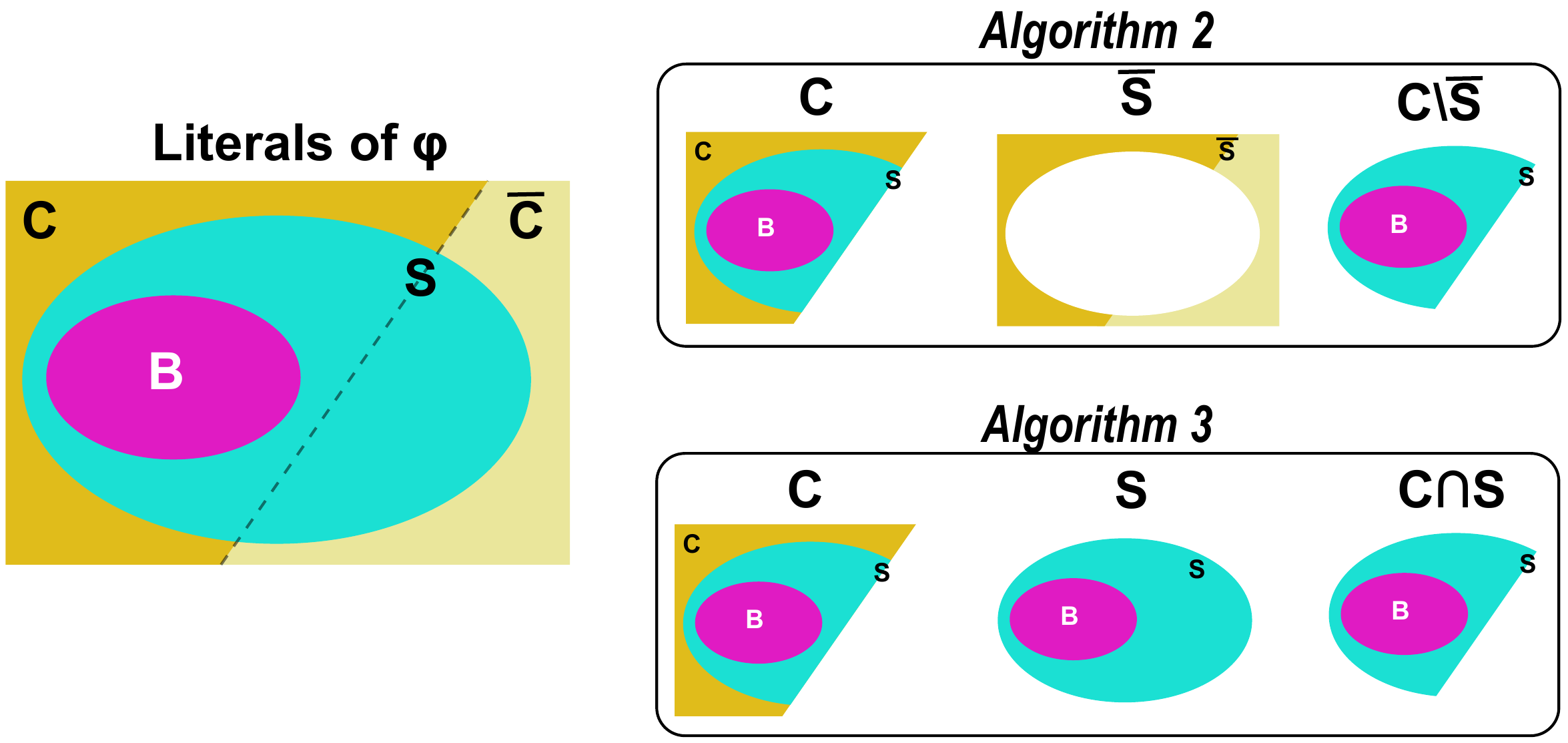}
     \caption{Graphical proof of $C \setminus \overline{S} = C \cap S$.}
     \label{fig:set-equivalence}
\end{figure*}

(i) the candidate set $C$ of literals that may be part of the backbone, and (ii) the remaining literals, denoted as $\overline{C}$. A solution $S$ is depicted in blue, with the backbone $B$ shown in pink (remember that $B \subseteq S$). On the right, the operations performed by Algorithms 2 and 3 are depicted step by step, demonstrating that both reach the same result.

\begin{algorithm}[htbp!]
\caption{\tool{Simplified Iterative}}
\begin{footnotesize}
\begin{flushleft}
\hspace*{\algorithmicindent} \textbf{Input}: A satisfiable formula $\varphi$ \\
\hspace*{\algorithmicindent} \textbf{Output}: The backbone $B$ of $\varphi$
\end{flushleft}
\begin{algorithmic}[1]
\State $B \gets \emptyset$
\State $(\mathrm{out}, C) \gets \mathrm{SAT}\big{(}\varphi\big{)}$  {\color{blue}\Comment{$B \subseteq C$}}
\State \textcolor{red}{$C \gets \mathrm{filter}(C)$}  {\color{blue}\Comment{Filter $C$}}

\While{$C \not= \emptyset$}
    \State $l \gets$ pick a literal from $C$
    \State $(\mathrm{out}, S) \gets \mathrm{SAT}\big{(}\varphi \cup \{\overline{l}\}\big{)}$   {\color{blue}\Comment{Is $l$ in the backbone?}}
    \If {$\mathrm{out}$ = \unsat{}}  {\color{blue}\Comment{Yes}}
        \State $B \gets B \cup \{l\}$
        \State $C \gets C \setminus \{l\}$
        \State{\color{blue}$\triangleright$ Help the solver by adding $l$ as unit clause}
        \State \textcolor{red}{$\varphi \gets \varphi \cup \{l\}$}
    \Else  {\color{blue}\Comment{No}}
        \State \textcolor{red}{$S \gets \mathrm{filter}(S)$}  {\color{blue}\Comment{Filter $S$}}
        \State $C \gets C \cap S$   {\color{blue}\Comment{$B = {\bigcap}_{\forall S}S$}}
    \EndIf
\EndWhile
\State  \Return{$B$}
\end{algorithmic}
\end{footnotesize}
\label{alg:simplified-iterative}
\end{algorithm}


The red code in Algorithm \ref{alg:simplified-iterative} implements two optional refinements, which can be incorporated into Algorithms \ref{alg:all-in} and \ref{alg:all-out} as well:

\begin{enumerate}[leftmargin=*]
    \item \textbf{Helping the solver}. When a literal is identified as part of the backbone, it can be added to the formula as a \textit{Unit Clause} (UC) (Line 11), thus aiding the solver in the subsequent calls \cite{Janota2015_Algorithms}. This can be done explicitly using the \textit{add} input of incremental solvers (see Fig. \ref{fig:sat-solver-inputs-outputs}). Additionally, some solvers, such as \tool{CaDiCaL}, perform this internally via a mechanism known as \textit{Transparent Incremental inProcessing} (TIP) \cite{Fazekas19_incremental_inprocessing}.

    \item \textbf{Filtering literals in solutions}. The solver's first solution establishes the initial backbone candidate set $C$ (Line 2). The subsequent solutions remove non-backbone literals from $C$ (Line 14). Some strategies can identify literals in the solutions that are or are not part of the backbone by just inspecting $\varphi$ superficially, thus reducing $C$ even more without wasting any solver call (Lines 3 and 13).

     Here, we present one of the most popular strategies, the \textit{rotatable literal filtering} devised by Janota et al. \cite{Janota2015_Algorithms}. Consider the solution $S = \{\overline{a}, \overline{b}, c, d, e, \overline{f}, \overline{g}\}$ for the formula in Table \ref{tab:cnf-example}. If we change one of the assignments in $\{\overline{a}, c, d, e, \overline{g}\}$, then $S$ will no longer be a valid solution. However, if we replace $\overline{b}$ with $b$ or $\overline{f}$ with $f$, then $S$ remains a solution. While the former literals might be part of the backbone, we can be certain that the latter literals, which are called \textit{rotatable} \cite{Janota2015_Algorithms} or \textit{flippable} \cite{Biere2023_CadiBack}, are not in the backbone and so they can be safely removed from $C$.

     Janota et al. \cite{Janota2015_Algorithms} proposed an algorithm that identifies \emph{Rotatable Literals} (RLs) in polynomial time relative to the number of clauses. Given a solution $S$, the algorithm iterates over each clause $\omega_i$ of $\varphi$ to compute the intersection $S \cap \omega_i$. If this intersection results in a single literal $\{l\}$, then that literal is considered \textit{unit}. After traversing all the clauses, a literal $l$ in $S$ is classified as rotatable if neither $l$ nor its negation $\overline{l}$ is unit.

     Zhang et al. \cite{Zhang2020_EDUCIBone} presented a tool called \tool{EDUCIBone}, which implements Algorithm \ref{alg:simplified-iterative} with three literal filtering strategies called \textit{COV}, \textit{WHIT} and \textit{2LEN}. Although the source code for \tool{EDUCIBone} is not publicly available, we will assess its performance on variability models in Section \ref{sec:experimental-evaluation}, as the binary code can be downloaded from \url{https://github.com/z971586668/DUCIBone/}.
\end{enumerate}

\subsubsection{Betting that All Literals in a Chunk Are Part of the Backbone}

Algorithms \ref{alg:naive-iterative}-\ref{alg:simplified-iterative} consume one solver call to test if one single literal is in the backbone or not. Janota et al. \cite{Janota2015_Algorithms} realized two cases where one call may answer a question involving multiple literals. Let $C'$ be a \textit{chunk} of literals that is a subset of $C$ (the set of literals candidate to be in the backbone):

\begin{itemize}[leftmargin=*]
    \item \textbf{Case 1: \textit{All-In}}.  $\mathrm{SAT}(\varphi \cup \overline{C'})$ returns \unsat{} if and only if all literals in $C'$ are in the backbone.\\
    \textit{Proof}: If there was a \sat{} assignment in $\varphi \wedge \overline{l_1}$ or in $\varphi \wedge \overline{l_2}$ $\ldots$ or in $\varphi \wedge \overline{l_n}$, then the corresponding(s) $l_i$('s) would be out of the backbone. Otherwise, no $l_i$'s are out of the backbone, which is equivalent to saying that all $l_i$'s are in the backbone. Due to the distributive law, $(\varphi \wedge \overline{l_1}) \vee  (\varphi \wedge \overline{l_2}) \vee \ldots \vee  (\varphi \wedge \overline{l_n}) \equiv \varphi \wedge (\overline{l_1} \vee \overline{l_2} \vee \ldots \vee \overline{l_n})$. Accordingly, $\mathrm{SAT}(\varphi \wedge \overline{l_1}) \vee \mathrm{SAT}(\varphi \wedge \overline{l_2}) \vee \ldots \vee \mathrm{SAT}(\varphi \wedge \overline{l_n})$ can be reformulated as $\mathrm{SAT}\Big{(}\varphi \wedge (\overline{l_1} \vee \overline{l_2} \vee \ldots \vee \overline{l_n})\Big{)} \equiv \mathrm{SAT}(\varphi \cup \overline{C'})$.
    \item \textbf{Case 2: \textit{All-Out}}. $\mathrm{SAT}\Big{(}\varphi \cup \{\overline{l} | l \in C'\}\Big{)}$ returns \sat{} if and only if all literals in $C'$ are out of the backbone.\\
    \textit{Proof}: $l_1, l_2, \ldots, l_n$ are all out of the backbone if $\varphi \wedge \overline{l_1}$ is \sat{}, and $\varphi \wedge \overline{l_2}$ is \sat{}, $\ldots$, and $\varphi \wedge \overline{l_n}$ is \sat{}. Due to the distributive law, $(\varphi \wedge \overline{l_1}) \wedge (\varphi \wedge \overline{l_2}) \wedge \ldots \wedge  (\varphi \wedge \overline{l_n}) \equiv \varphi \wedge (\overline{l_1} \wedge \overline{l_2} \wedge \ldots \wedge \overline{l_n})$. Accordingly, $\mathrm{SAT}(\varphi \wedge \overline{l_1}) \wedge  \mathrm{SAT}(\varphi \wedge \overline{l_2}) \wedge \ldots \wedge  \mathrm{SAT}(\varphi \wedge \overline{l_n})$ can be reformulated as $\mathrm{SAT}\Big{(}\varphi \wedge (\overline{l_1} \wedge \overline{l_2} \wedge \ldots \wedge \overline{l_n})\Big{)} \equiv \mathrm{SAT}\Big{(}\varphi \cup \{\overline{l} | l \in C'\}\Big{)}$.
\end{itemize}

Algorithms \ref{alg:all-in} and \ref{alg:all-out} exploit Cases 1 and 2, respectively\footnote{In \cite{Janota2015_Algorithms} and \cite{Biere2023_CadiBack},  Algorithms \ref{alg:all-in} and \ref{alg:all-out} are called \textit{Chunking Algorithm} and \textit{Core-based Algorithm with Chunking}, respectively. We have changed their names to \textit{Bet All-In} and \textit{Bet All-Out}, as we think they reflect the algorithms' spirit better.}. 

\begin{algorithm}
\caption{\tool{Bet All-In}}
\begin{footnotesize}
\begin{flushleft}
\hspace*{\algorithmicindent} \textbf{Input}: A satisfiable formula $\varphi$; the literals' chunk size $k$ \\
\hspace*{\algorithmicindent} \textbf{Output}: The backbone $B$ of $\varphi$
\end{flushleft}
\begin{algorithmic}[1]
\State $B \gets \emptyset$
\State $(\mathrm{out}, C) \gets \mathrm{SAT}\big{(}\varphi\big{)}$
\While { $C \neq \emptyset$ }
    \State{\color{blue}$\triangleright$ Get a chunk $C'$ from the candidate set $C$}
    \State $k' \gets \mathrm{min}(k, |C|)$
    \State $C' \gets$ pick $k'$ literals from $C$
    \State $(\mathrm{out}, S) \gets 
    \mathrm{SAT}(\varphi \cup \overline{C'})$
    \If {$\mathrm{out}$ = \unsat{}}
        \State{\color{blue}$\triangleright$ We win: all literals in $C'$ are in the backbone}
        \State $B \gets B \cup C'$
        \State $C \gets C \setminus C'$
    \Else
        \State{\color{blue} $\triangleright$ We lose: at least one literal in $C'$ isn't}
        \State{\color{blue} in the backbone}
        \State $C \gets C \cap S$
    \EndIf
\EndWhile
\State  \Return{$B$}
\end{algorithmic}
\end{footnotesize}
\label{alg:all-in}
\end{algorithm}

It is important to note that:

\begin{itemize}[leftmargin=*]
    \item The effectiveness of Algorithms \ref{alg:all-in} and \ref{alg:all-out} relies heavily on how frequently Cases 1 and 2 occur. For example, Algorithm \ref{alg:all-in} will outperform Algorithm \ref{alg:simplified-iterative} in variability models where most literals are part of the backbone and the chunk size is adjusted to trigger Case 1 often. Otherwise, Algorithm \ref{alg:simplified-iterative} will be faster.
    \item When the chunk size is $k=1$, Algorithms \ref{alg:all-in} and \ref{alg:all-out} emulate Algorithm~\ref{alg:simplified-iterative} behavior.
\end{itemize}

As represented in Fig. \ref{fig:sat-solver-inputs-outputs}, incremental SAT solvers support assuming literals, but not clauses, which, in contrast, can be permanently \emph{added}. To overcome this limitation, \tool{MiniBones} \cite{Janota2015_Algorithms} implements Line 7 of Algorithm \ref{alg:all-in} with a technique called \textit{activation literals}, described in the pages 153-154 of \cite{handbook}, which works as follows:
\begin{enumerate}[leftmargin=*]
    \item A \textit{fresh} activation literal $a$ is created for each clause $\omega$ that will be turned on/off in Line 7 of Algorithm \ref{alg:all-in} ($\omega = \vee_{l \in C'}\overline{l}$ and $a$ is a new literal not present anywhere in the formula).
    \item The clauses $\omega \vee a$ are added to the formula before any SAT call is performed.
    \item Then, to turn $\omega$ on, the solver is called assuming the literal $\overline{a}$.
    \item To turn $\omega$ off, the solver is called assuming the literal $a$.
\end{enumerate}

Remember that, as pointed out in Section \ref{sec:sat-solver}, when the maximum number of literals per clause is limited to 2, the SAT problem can be solved in polynomial time. However, as the number of literals increases, the problem becomes significantly more challenging. Accordingly, the drawback of the \textit{activation literals} technique is that every time it is applied, a new clause containing $|C'| + 1$ literals is added to the formula, thus progressively decreasing the SAT solver performance \cite{Yu14}.

\subsubsection{Betting that No Literals in a Chunk Are Part of the Backbone}

Algorithm \ref{alg:all-out} bets that no literals in the chunk are in the backbone, i.e., it exploits situations where most literals are not in the backbone and the chunk size is adjusted to trigger Case 2. When this fails, Algorithm \ref{alg:all-out} uses a last resort: the UNSAT core $U$ (one of the outputs depicted in Fig. \ref{fig:sat-solver-inputs-outputs}).

$U$ includes assumptions that makes $\varphi$ \unsat{}. However, remember that current solvers do not guarantee minimality, which means that $U$ may include assumptions that are not responsible for the conflict. Nevertheless, in the extreme case where $U \cap \{\overline{l} | l \in C'\}$ is a single literal $m$, minimality is ensured (there is no alternative possibility!), and so we can conclude that $\overline{m}$ is in the backbone (Lines 16-20).

Algorithm \ref{alg:all-out} keeps playing the betting game until $\{\overline{l} | l \in C'\}$ is exhausted with successive UNSAT cores (Line 22). At that point, the remaining literals in the chunk are tested following Algorithm \ref{alg:simplified-iterative} style (Lines 24-34).

\begin{algorithm}
\caption{\tool{Bet All-Out}}
\begin{footnotesize}
\begin{flushleft}
\hspace*{\algorithmicindent} \textbf{Input}: A satisfiable formula $\varphi$; the literals' chunk size $k$ \\
\hspace*{\algorithmicindent} \textbf{Output}: The backbone $B$ of $\varphi$
\end{flushleft}
\begin{algorithmic}[1]
\State $B \gets \emptyset$
\State $(\mathrm{out}, C, U) \gets \mathrm{SAT}\big{(}\varphi\big{)}$
\While { $C \neq \emptyset$ }
    \State{\color{blue}$\triangleright$ Process one chunk of literals}
    \State $k' \gets \mathrm{min}(k, |C|)$
    \State $C' \gets$ pick $k'$ literals from $C$  {\color{blue}\Comment $C'$ is the chunk}
    \While {true}
        \State $(\mathrm{out}, S, U) \gets 
        \mathrm{SAT}\Big{(}\varphi \cup \{\overline{l} | l \in C'\}\Big{)}$    
        \If {$\mathrm{out}$ = \sat{}}
            \State{\color{blue}$\triangleright$ We win: no literals in $C'$ are in the backbone}
            \State $C \gets C \cap S$
            \State $\textbf{break}$  {\color{blue}\Comment{Go for the next chunk}}
        \Else
        \State{\color{blue}$\triangleright$ We lose: at least one literal in $C'$}
        \State{\color{blue}is in the backbone}
            \If {$U \cap \{\overline{l} | l \in C'\} = \{m\}$}
                \State{\color{blue}$\triangleright$ We recovered: $\overline{m}$ is in the backbone}
                \State $B \gets B \cup \{ \overline{m} \}$
                \State $C' \gets C' \setminus \{ \overline{m}\}$
                \State $C \gets C  \backslash \{ \overline{m} \}$
            \EndIf
            \State{\color{blue} $\triangleright$ Keep testing the literals $C'$ that are not in $U$}
            \If {$\{\overline{l} | l \in C'\} \setminus U = \emptyset$}
                \State{\color{blue} $\triangleright$ The UNSAT core cannot help advance}
                \While{$C' \not= \emptyset$}
                    \State{\color{blue} $\triangleright$ Test literals in $C'$ as Algorithm \ref{alg:simplified-iterative}}
                    \State $l \gets$ pick a literal from $C'$
                    \State $(\mathrm{out}, S, U) \gets \mathrm{SAT}\big{(}\varphi \cup \{\overline{l}\}\big{)}$
                    \If {$\mathrm{out}$ = \unsat{}}
                        \State $B \gets B \cup \{l\}$
                        \State $C' \gets C' \setminus \{l\}$
                        \State $C \gets C \setminus \{l\}$
                    \Else
                        \State $C' \gets C' \cap S$
                        \State $C \gets C \cap S$
                    \EndIf
                \EndWhile
                \State $\textbf{break}$  {\color{blue}\Comment{Go for the next chunk}}
            \EndIf
        \EndIf
    \EndWhile
\EndWhile
\State  \Return{$B$}
\end{algorithmic}
\end{footnotesize}
\label{alg:all-out}
\end{algorithm}

\subsection{Tools and Conflicting Findings in Published Experiments}\label{sec:tools}

Table \ref{tab:tool-summary} summarizes the tools this paper examines, indicating the algorithms and features they support:

\begin{table*}[htpb!]
\caption{Algorithms and features provided by SPL tools (\tool{FlamaPy} and \tool{FeatureIDE}) and backbone solvers from the SAT community (\tool{MiniBones}, \tool{CadiBack} and \tool{EDUCIBone}).}
\label{tab:tool-summary}
\begin{center}
\begin{scriptsize}
\begin{tabular}{|c|c|c|c|c|c|c|}
\hline
    \multirow{2}{*}{\textbf{Tool}} &
    \multirow{2}{*}{\textbf{Alg. \ref{alg:naive-iterative}}} &
    \multirow{2}{*}{\textbf{Alg. \ref{alg:advanced-iterative}/\ref{alg:simplified-iterative}}} &
    \multirow{2}{*}{\textbf{Alg. \ref{alg:all-in}}} &
    \multirow{2}{*}{\textbf{Alg. \ref{alg:all-out}}} &
    \textbf{Helps} &
    \textbf{Filters} \\
    &
    &
    &
    &
    &
    \textbf{solver?} &
    \textbf{literals?} \\ \hline \hline
    \tool{FlamaPy}  & \cmark & & & & & \\ \hline
    \tool{FeatureIDE} & & \cmark & & & & \\ \hline \hline
    \multirow{2}{*}{\tool{MiniBones}} & & Emulated & \multirow{2}{*}{\cmark} & \multirow{2}{*}{\cmark} & Unit  &  Rotating\\
    & & with $k = 1$ & & & clause & literals\\ \hline
    \multirow{2}{*}{\tool{CadiBack}} & & Emulated & \multirow{2}{*}{\cmark} & \multirow{2}{*}{\cmark} & Transparent inc. & Rotating \\
    & & with $k = 1$ & & & processing & literals\\ \hline
    \multirow{2}{*}{\tool{EDUCIBone}}& & \multirow{2}{*}{\cmark} &  &  & Unit & COV, WHIT  \\
    & & & & & clause & and 2LEN\\ \hline
\end{tabular}
\end{scriptsize}
\end{center}
\end{table*}

\begin{enumerate}[leftmargin=*]
    \item \tool{FlamaPy}\footnote{\url{https://www.flamapy.org/}} \cite{Galindo2023_Flama} and \tool{FeatureIDE}\footnote{\url{https://featureide.github.io/}} \cite{Thum2014-FeatureIDE}. Two of the most popular and updated tools for the automated analysis of variability models. \tool{FlamaPy} is written in \tool{Python}, and \tool{FeatureIDE} in \tool{Java}.
    \item Three state-of-the-art backbone extractors developed within the SAT community, all of them written in \tool{C++}:
    \begin{enumerate}[leftmargin=6pt]
        \item \tool{MiniBones}\footnote{\url{https://github.com/MikolasJanota/minibones}} \cite{Janota2015_Algorithms}. Released in 2015, it proposed Algorithms \ref{alg:all-in} and \ref{alg:all-out} and implemented them on top of the incremental SAT solver \tool{MiniSAT 2.2}. \tool{MiniBones}  represents a fundamental milestone and has served as the baseline for the subsequent backbone solvers.

        \item \tool{EDUCIBone}\footnote{\url{https://github.com/z971586668/DUCIBone/}} \cite{Zhang2020_EDUCIBone}. Released in 2020, it implements Algorithm \ref{alg:simplified-iterative} with three literal filtering mechanisms called \textit{COV}, \textit{WHIT} and \textit{2LEN}.
            
        \item \tool{CadiBack}\footnote{\url{https://github.com/arminbiere/cadiback}} \cite{Biere2023_CadiBack}. Released in 2024, it reimplements Algorithms \ref{alg:all-in} and \ref{alg:all-out} with the modern solver \tool{CaDiCaL}, (i) improving the rotatable literal filtering, (ii) seizing \tool{CaDiCaL}'s TIP (the advanced alternative to injecting backbone literals in the formulas as UCs mentioned in Section \ref{sec:filtering}), and (iii) using a dynamic chunk size that instead of being set by the user, it starts at $k=1$ and is multiplied by a constant $K$ with each loop iteration, provided that the all-in/all-out condition is satisfied; if the condition is not met, $k$ is reset to 1. In particular, \cite{Biere2023_CadiBack} explores three $K$ values for \tool{CadiBack}: $K = 1$ to behave as Algs. \ref{alg:advanced-iterative}/\ref{alg:simplified-iterative}, $K = 10$ to test \textit{chunks} of literals, and $K = \infty$ to consider a single chunk containing all the variables of the formula.
    \end{enumerate}
\end{enumerate}

The algorithms and features supported by \tool{MiniBones}, \tool{EduciBones}, and \tool{CadiBack} have been evaluated on formulas from SAT competitions in \cite{Janota2015_Algorithms,Biere2023_CadiBack,Zhang2020_EDUCIBone}. However, the reported results are contradictory:
\begin{enumerate}[leftmargin=*]
    \item According to Janota et al.'s experiments on 779 formulas \cite{Janota2015_Algorithms}:
    \begin{enumerate}[leftmargin=6pt]
        \item Algorithm \ref{alg:all-out} is the fastest whenever less than 25\% of the literals are in the backbone. Otherwise, there is no significant difference between Algorithms \ref{alg:all-in} and \ref{alg:all-out}.
        \item As $k$ increases, the performance of Alg. \ref{alg:all-out} improves until it reaches a size where the performance begins to decrease. The best performance was achieved at $k=100$, after which it decreased significantly, particularly from $k=500$.
        \item Testing four $k$ values on Alg. \ref{alg:all-in}  (1, 30, 100, and 500), the authors found that as $k$ increased, the number of SAT calls decreased, but the time per SAT call increased more significantly. Accordingly, they concluded that the performance of Algorithm \ref{alg:all-in} decreases as $k$ increases. Alg. \ref{alg:all-in} reached its worst performance when $k$ was equal to the total number of variables.
        \item Two literal filtering techniques were tested: rotating literals and set covering. Both of them were detrimental to Algs. \ref{alg:simplified-iterative}-\ref{alg:all-out}.
    \end{enumerate}
    Consequently, \tool{MiniBones}'s default configuration is Alg. \ref{alg:all-out} with $k = 100$ and without filtering rotatable literals.
    \item According to Biere et al.'s experiments on 1,798 formulas \cite{Biere2023_CadiBack}:
    \begin{enumerate}[leftmargin=6pt]
        \item Algorithm \ref{alg:all-in} is always the fastest.
        \item As $k$ increases, the performance of Algorithm \ref{alg:all-in} increases, reaching its highest when $k$ equals the total number of variables in the formula (i.e., when $K = \infty$).
        \item Filtering rotatable literals does not affect the performance of Algorithms \ref{alg:simplified-iterative}-\ref{alg:all-out}. 
    \end{enumerate}
    Consequently, \tool{CadiBack}'s default configuration is Alg. \ref{alg:all-in} with $K~=~\infty$ and filtering rotatable literals.
    \item According to Zhang et al.'s experiments on  1,816 formulas \cite{Zhang2020_EDUCIBone} :
    \begin{enumerate}[leftmargin=6pt]
        \item Algorithm \ref{alg:simplified-iterative} with the strategies COV, WHIT, and 2LEN is faster than Algorithm \ref{alg:all-out} with $k=100$.
        \item Filtering literals, particularly with COV, WHIT, and 2LEN, greatly improves the performance of Algorithm \ref{alg:simplified-iterative}.
    \end{enumerate}
    As mentioned before, the source code of \tool{EDUCIBone} is not available. Only the binary code is provided, with no accompanying documentation that could explain its default configuration. We will suppose \tool{EDUCIBone}'s default configuration is the one that gets the best results in \cite{Zhang2020_EDUCIBone}: Alg. \ref{alg:simplified-iterative} with the literal filtering strategies COV, WHIT, and 2LEN.

\end{enumerate}

\section{Empirical Evaluation}\label{sec:experimental-evaluation}

This section describes the experimental evaluation we conducted to answer the following \textit{Research Questions} (RQs):

\begin{itemize}
    \item Q1: Characterization of VM formulas. \textit{Are formulas obtained from VMs different from those used in SAT competitions?}
    \item Q2: Comparing the algorithms that SPL tools implement. \textit{Is Alg.~\ref{alg:advanced-iterative}/\ref{alg:simplified-iterative} faster than Alg.~\ref{alg:naive-iterative}?}
    \item Q3: Comparing SPL algorithms to SAT community algorithms. \textit{Are there faster algorithms than Alg.~\ref{alg:naive-iterative} or Alg.~\ref{alg:advanced-iterative}/\ref{alg:simplified-iterative}?}
    \item Q4: Estimation of the best $k$. \textit{Is it possible to infer the optimal $k$ for Algs. \ref{alg:all-in} and \ref{alg:all-out} from easily computable formula features, e.g., the number of variables or clauses?}
    \item Q5: Influence of UC injection. \textit{Does UC injection influence the algorithms’ performance?}
    \item Q6: Influence of TIP. \textit{Does TIP influence the algorithms’ performance?}
    \item Q7: Influence of RLs. \textit{Does literal filtering with RLs influence the algorithms’ performance?}
    \item Q8: Influence of COV, WHIT and 2LEN. \textit{Does the combined filtering with COV, WHIT, and 2LEN influence the algorithms’ performance?}
\end{itemize}

\subsection{Experimental Setup}

To answer all RQs, we used a benchmark with 2,371 formulas derived from real VMs by:

\begin{itemize}
\item Sundermann et al.~\cite{Sundermann24benchmark} using the \tool{TraVarT} translator\footnote{ \url{https://zenodo.org/records/11654486}~\cite{Feichtinger21}.}.
\item Fernandez-Amoros et al.~\cite{fernandez23} using the \tool{KconfigSampler} tool\footnote{\url{https://github.com/davidfa71/Sampling-the-Linux-kernel}~\cite{fernandez23}.}.
\end{itemize}

In addition, to answer RQ1, we also examined the structure of the 1,798 formulas from the SAT competition benchmarks (2004-2022) used in \cite{Biere2023_CadiBack} to assess the performance of \tool{CadiBack} and \tool{MiniBones}.

\tool{MiniBones}, \tool{CadiBack}, and \tool{EDUCIBone} are implemented in C++, while \tool{FlamaPy} is implemented in Python and \tool{FeatureIDE} in Java. To ensure fairness and minimize bias related to programming language differences, Algs.~\ref{alg:naive-iterative} and \ref{alg:advanced-iterative}/\ref{alg:simplified-iterative} were implemented in C++ using the IPASIR interface. The resulting tool, \tool{IPASIRBones}, is available at \url{https://github.com/lcambelo/ipasirbones}. Since \tool{IPASIRBones} invokes the underlying SAT-solver via IPASIR, users may select \tool{MiniSat}, \tool{CaDiCaL}, or use any other solver compatible with the IPASIR interface.

To perform our study, we tested the configurations summarized in Table \ref{tab:config-summary}.

\begin{table*}[htpb!]
\caption{Configurations tested in the experiments. (*) is the default configuration for \tool{CadiBack} and (**) with $k=100$ the default for \tool{MiniBones}.}
\label{tab:config-summary}
\begin{center}
\begin{scriptsize}
\begin{tabular}{|c|c|c|c|c|c|c||c|c|}
\hline
    \multirow{2}{*}{\textbf{Tool}} &
    \textbf{Alg.} &
    \textbf{Alg.} &
    \textbf{Alg.} &
    \textbf{Alg.} &
    \textbf{Helps} &
    \textbf{Filters} & \multirow{2}{*}{\textbf{Id}} &
    \multirow{2}{*}{\textbf{\#Executions}} \\
    & \textbf{\ref{alg:naive-iterative}}
    & \textbf{\ref{alg:advanced-iterative}/\ref{alg:simplified-iterative}}
    & \textbf{\ref{alg:all-in}}
    & \textbf{\ref{alg:all-out}}
    &
    \textbf{solver?} &
    \textbf{literals?} & & \\ \hline \hline
    \tool{IPASIRBones}  & \multirow{2}{*}{\cmark} & & & & & & \multirow{2}{*}{C1} &  \multirow{2}{*}{7,113} \\
    \tool{\& MiniSat}  & & & & & & & &  \\ \hline \hline
    \tool{IPASIRBones} & & \multirow{2}{*}{\cmark} & & & & & \multirow{2}{*}{C2} & \multirow{2}{*}{7,113}\\
    \tool{\& MiniSat}  & & & & & & & &  \\ \hline
    \tool{IPASIRBones} & & \multirow{2}{*}{\cmark} & & & \multirow{2}{*}{UC} & & \multirow{2}{*}{C3} & \multirow{2}{*}{7,113}\\
    \tool{\& MiniSat}  & & & & & & & &  \\ \hline
    \tool{IPASIRBones} & & \multirow{2}{*}{\cmark} & & & \multirow{2}{*}{TIP} & & \multirow{2}{*}{C4} & \multirow{2}{*}{7,113}\\
    \tool{\& CaDiCaL}  & & & & & & & &  \\ \hline
    \tool{MiniBones} & & \multirow{2}{*}{\cmark}& & & & \multirow{2}{*}{RL} &  \multirow{2}{*}{C5} & \multirow{2}{*}{7,113}\\
    \tool{$k=1$}  & & & & & & & &  \\ \hline
    \tool{MiniBones} & & \multirow{2}{*}{\cmark} & & & \multirow{2}{*}{UC} & \multirow{2}{*}{RL} & \multirow{2}{*}{C6} & \multirow{2}{*}{7,113}\\
    \tool{$k=1$}  & & & & & & & &  \\ \hline
    \tool{CadiBack} & & \multirow{2}{*}{\cmark} & & & \multirow{2}{*}{TIP} & \multirow{2}{*}{RL} & \multirow{2}{*}{C7} & \multirow{2}{*}{7,113}\\
    \tool{$k=1$}  & & & & & & & &  \\ \hline \hline
    \tool{MiniBones}  & &  & \multirow{2}{*}{\cmark} &  &  &  & \multirow{2}{*}{C8} & \multirow{2}{*}{149,373}\\
    \tool{$k= 5, \ldots, 100, \#\mathrm{vars}$}  & & & & & & & &  \\ \hline
    \tool{MiniBones} & &  & \multirow{2}{*}{\cmark} &  &  \multirow{2}{*}{UC} & & \multirow{2}{*}{C9} & \multirow{2}{*}{149,373}\\
    \tool{$k= 5, \ldots, 100, \#\mathrm{vars}$}  & & & & & & & &  \\ \hline
    \tool{CadiBack} & & & \multirow{2}{*}{\cmark} & & \multirow{2}{*}{TIP} & & \multirow{2}{*}{C10} & \multirow{2}{*}{7,113}\\
    \tool{$k=\mathrm{aut}$}  & & & & & & & &  \\ \hline
    \tool{MiniBones} & &  & \multirow{2}{*}{\cmark} &  &  &  \multirow{2}{*}{RL} & \multirow{2}{*}{C11} &\multirow{2}{*}{149,373}\\
    \tool{$k= 5, \ldots, 100, \#\mathrm{vars}$}  & & & & & & & &  \\ \hline
    \tool{MiniBones} & &  & \multirow{2}{*}{\cmark} &  &  \multirow{2}{*}{UC} & \multirow{2}{*}{RL} & \multirow{2}{*}{C12} & \multirow{2}{*}{149,373}\\
    \tool{$k= 5, \ldots, 100, \#\mathrm{vars}$}  & & & & & & & &  \\ \hline
    \tool{CadiBack} & & & \multirow{2}{*}{\cmark} & & \multirow{2}{*}{TIP} & \multirow{2}{*}{RL} & \multirow{2}{*}{C13} & \multirow{2}{*}{7,113}\\
    \tool{$k=\mathrm{aut}$}  & & & & & & & &  \\ \hline
    \tool{CadiBack} & & & \multirow{2}{*}{\cmark} & & \multirow{2}{*}{TIP} & \multirow{2}{*}{RL} & \multirow{2}{*}{C14$^*$} & \multirow{2}{*}{7,113}\\
    \tool{$k=\infty$}  & & & & & & & &  \\ \hline \hline

    \tool{MiniBones}  & &  & & \multirow{2}{*}{\cmark}  &  &  & \multirow{2}{*}{C15} & \multirow{2}{*}{149,373}\\
    \tool{$k= 5, \ldots, 100, \#\mathrm{vars}$}  & & & & & & & &  \\ \hline
    \tool{MiniBones} & &  & & \multirow{2}{*}{\cmark}  &  \multirow{2}{*}{UC} & & \multirow{2}{*}{C16$^{**}$} & \multirow{2}{*}{149,373}\\
    \tool{$k= 5, \ldots, 100, \#\mathrm{vars}$}  & & & & & & & &  \\ \hline
    \tool{CadiBack} & & & & \multirow{2}{*}{\cmark} & \multirow{2}{*}{TIP} & & \multirow{2}{*}{C17} & \multirow{2}{*}{7,113} \\
    \tool{$k=\mathrm{aut}$}  & & & & & & & &  \\ \hline
    \tool{MiniBones} & &  & & \multirow{2}{*}{\cmark}  & & \multirow{2}{*}{RL} & \multirow{2}{*}{C18} & \multirow{2}{*}{149,373}\\
    \tool{$k= 5, \ldots, 100, \#\mathrm{vars}$}  & & & & & & & &  \\ \hline
    \tool{MiniBones} & &  & & \multirow{2}{*}{\cmark}  &  \multirow{2}{*}{UC} & \multirow{2}{*}{RL} & \multirow{2}{*}{C19} & \multirow{2}{*}{149,373}\\
    \tool{$k= 5, \ldots, 100, \#\mathrm{vars}$}  & & & & & & & &  \\ \hline
    \tool{CadiBack} & & & & \multirow{2}{*}{\cmark} & \multirow{2}{*}{TIP} & \multirow{2}{*}{RL} & \multirow{2}{*}{C20} & \multirow{2}{*}{7,113}\\
    \tool{$k=\mathrm{aut}$}  & & & & & & & &  \\ \hline \hline

    \multirow{2}{*}{\tool{EDUCIBone}}& & \multirow{2}{*}{\cmark} &  &  & \multirow{2}{*}{UC} & COV, WHIT & \multirow{2}{*}{C21} & \multirow{2}{*}{7,113}\\
    & & & & & & \& 2LEN & &\\ \hline \hline
    \multicolumn{8}{|r}{Total executions:} & 1,287,453\\ \hline
\end{tabular}
\end{scriptsize}
\end{center}
\end{table*}

\newpage

Each configuration-$k$ combination in the table was run three times on an HP ProLiant server with an Xeon E5-2660 v4 processor. A total of 1,287,453 executions were performed, requiring 59.49 days of computation time. Subsequently, the median times for the three-group runs were further analyzed.

The data characterizing the VMs, including VM name, number of variables and clauses, median number of literals per clause, and number of clauses with more than two literals, as well as execution details such as running time, configuration identifiers, and command line parameters for the backbone tool, along with the code required to replicate our experimental analysis, are available at the following Zenodo repository: \\
\url{https://doi.org/10.5281/zenodo.18494650}

\begin{figure*}[htbp!]
    \centering
    \includegraphics[width=0.8\linewidth]{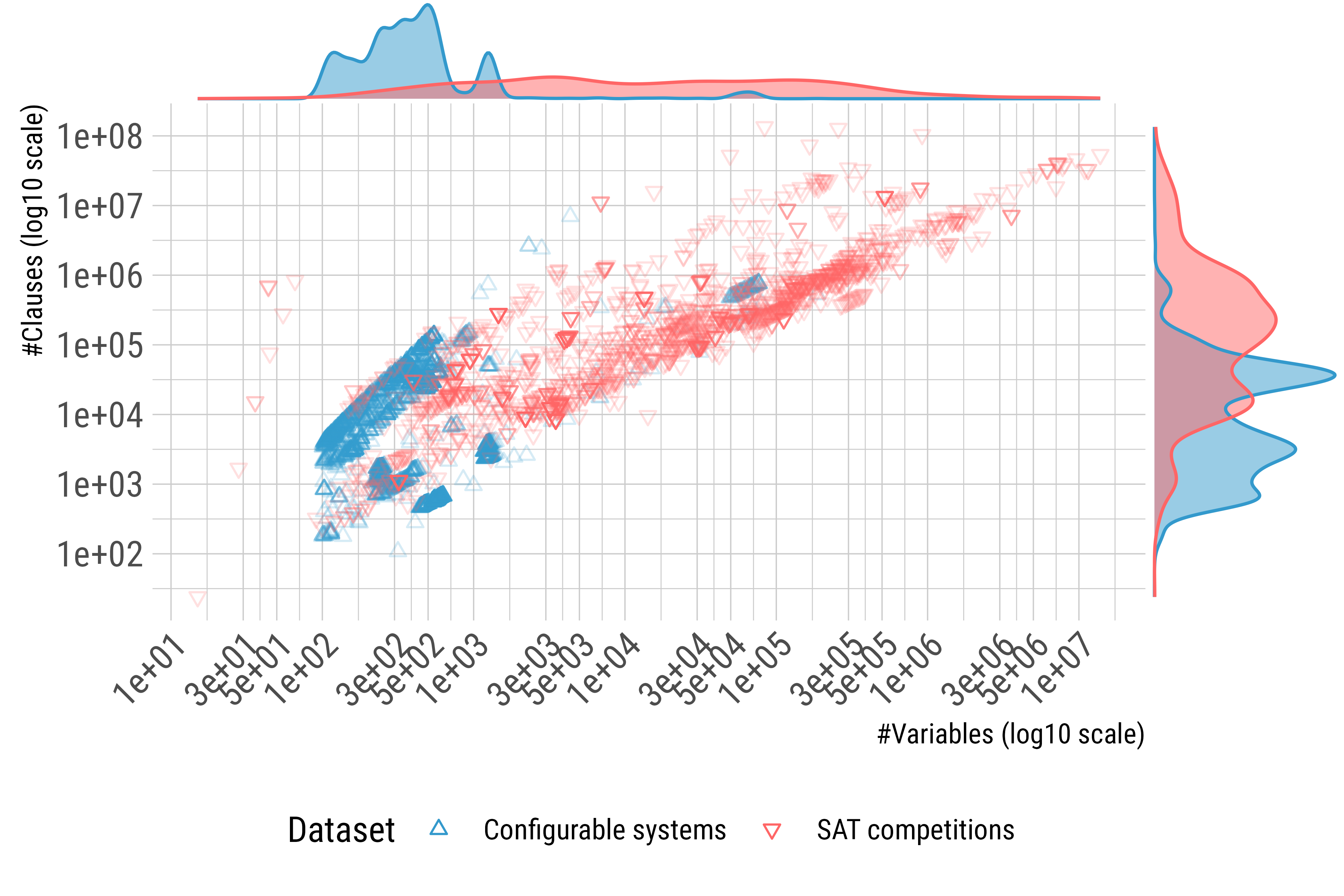}
    \caption{Scatter plot illustrating the complexity of the formulas in the VM and SAT competition datasets. Each point represents a formula, with the $x$- and $y$-coordinates corresponding to the number of variables and clauses, respectively. The density diagrams at the top and right display the marginal distributions of the number of clauses and variables, respectively.}\label{fig:nclauses-vs-nvars}
\end{figure*}

\begin{table*}[htbp!]
\caption{Complexity of VM formulas compared to SAT competition formulas from \cite{Biere2023_CadiBack}. The reported values are medians computed across all formulas in each dataset.}
\label{tab:datasets-summary}
\centering
\begin{tabular}{|l|r|r|r|r|}
  \hline
\textbf{Dataset} & \textbf{\#vars}& \textbf{$\frac{\mathrm{\#clauses}}{\mathrm{\#vars}}$} & \textbf{\#literals} & \textbf{\%clauses with} \\
 & &  & \textbf{per clause} & \textbf{more than 2 literals} \\
  \hline \hline
Configurable systems & 356 & 39.84 & 2 & 2.01\% \\ \hline
SAT competitions & 13,408 & 5.46 & 3 & 61.32\% \\ \hline
\end{tabular}
\end{table*}

\subsection{Results}

This section summarizes the experimental results for each research question.

\subsubsection{Q1: Characterization of VM Formulas}\label{sec:q1}

Existing experimental evaluations of backbone algorithms rely on formulas from SAT competitions, with the most comprehensive benchmark being compiled in \cite{Biere2023_CadiBack}. Fig. \ref{fig:nclauses-vs-nvars} and Table \ref{tab:datasets-summary} compare the complexity of the VM formulas in our dataset (in blue) against the formulas in \cite{Biere2023_CadiBack} (in red). 

All columns in Table \ref{tab:datasets-summary} display median values; for example, the first column shows that VM formulas have a median of 356 variables, whereas SAT competition formulas have a median of 13,408 variables. The column labeled ``\#literals per clause'' reports a double median: for each formula, the median number of literals per clause is calculated, and the column then shows the median of these values across all formulas in each dataset. The following conclusions can be drawn from Fig. \ref{fig:nclauses-vs-nvars} and Table \ref{tab:datasets-summary}:

\begin{itemize}[leftmargin=*]
    \item VM formulas have fewer variables.
    \item In proportion, VM formulas have many more clauses.
    \item Clauses of VM formulas are much simpler, typically with just 2 literals (remember from Section \ref{sec:sat-solver} that the 2-SAT problem can be efficiently solved in polynomial time).
\end{itemize}

\begin{center}
\begin{tcolorbox}[summary,title=\textbf{Answer to Q1: \textit{Are formulas obtained from VMs different from those used in SAT competitions?}}]
\textbf{Yes, they are.}
VM formulas differ substantially in the number of variables, clause density, and clause complexity: they have fewer variables and more clauses, but about 98\% of these clauses contain at most two literals.
\end{tcolorbox}
\end{center}

\subsubsection{Q2: Comparing the Algorithms that SPL Tools Implement}\label{sec:q2}

\begin{figure*}[htbp!]
    \centering
    \includegraphics[width=1\linewidth]{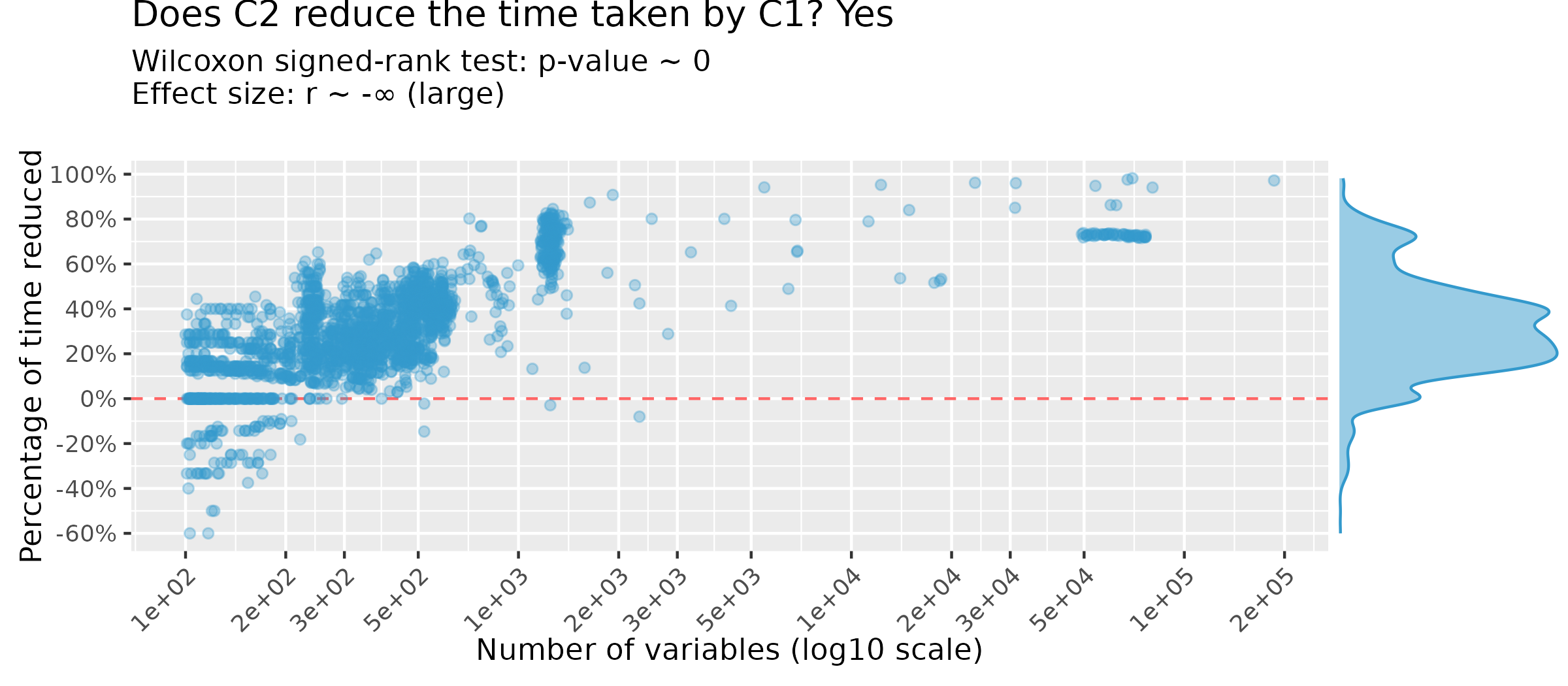}
    \caption{Scatter plot showing the percentage reduction in C1’s runtime achieved by C2. Each point represents a formula, with its x-coordinate denoting the formula’s number of variables and its y-coordinate the percentage reduction in runtime achieved by C2 over C1. The dashed red line indicates no reduction; points above it indicate formulas where C2 outperformed C1, while points below it indicate C2 was slower. The density diagram on the right displays the marginal distribution of percentage time reduction.}\label{fig:c1_c2}
\end{figure*}

Fig. \ref{fig:c1_c2} shows that C2 (Alg. \ref{alg:advanced-iterative}/\ref{alg:simplified-iterative}) is faster than C1 (Alg. \ref{alg:naive-iterative}) for the practical totality of VM formulas, regardless of their number of variables. Each point corresponds to a formula. The $x$-axis represents the number of variables in the formula, and the $y$-axis denotes the percentage reduction in runtime achieved by C2 relative to C1 calculated as follows: let $t_{\mathrm{C1}}$ and $t_{\mathrm{C2}}$ denote the times required by C1 and C2 to compute the backbone, respectively; then, the percentage reduction is given by $\frac{t_{\mathrm{C1}}-t_{\mathrm{C2}}}{t_{\mathrm{C1}}}\cdot 100$.

\vspace{0.25cm}
The scatterplot helps visualize the algorithms' performance depending on the formula size. However, there is overlap among data points that complicates the interpretation of their distribution. The right density plot clarifies this distribution. The red line represents no reduction; areas above the line indicate a positive reduction, while areas below the line indicate a negative reduction (i.e., an increase in time). In the density plot, most of the area lies above the red dashed line. Accordingly, C2 reduces the time required by C1 to obtain the backbone in the majority of the VM formulas.

Fig. \ref{fig:c1_c2}  also displays the results of a Wilcoxon signed-rank test comparing C2 times to C1 times across all formulas. The result shows the difference is statistically significant (the $p$-value is practically zero) and large (according to the guidelines in \cite{Cohen88}, $|r|<0.3$ should be considered a small effect size, $0.30\leq |r| <0.5$ medium, and $|r|>0.5$ large).

\begin{center}
\begin{tcolorbox}[summary,title=\textbf{Answer to Q2: \textit{Is Alg.~2/3 faster than Alg.~1?}}]
\textbf{Yes, it is.}
Alg.~\ref{alg:advanced-iterative}/\ref{alg:simplified-iterative} is consistently faster than Alg.~\ref{alg:naive-iterative} across the entire dataset, independently of the formulas’ number of variables.
\end{tcolorbox}
\end{center}

\subsubsection{Q3: Comparing SPL Algorithms to SAT Community Algorithms}\label{sec:q3}

\begin{figure*}[htbp!]
    \centering
    \includegraphics[width=0.95\linewidth]{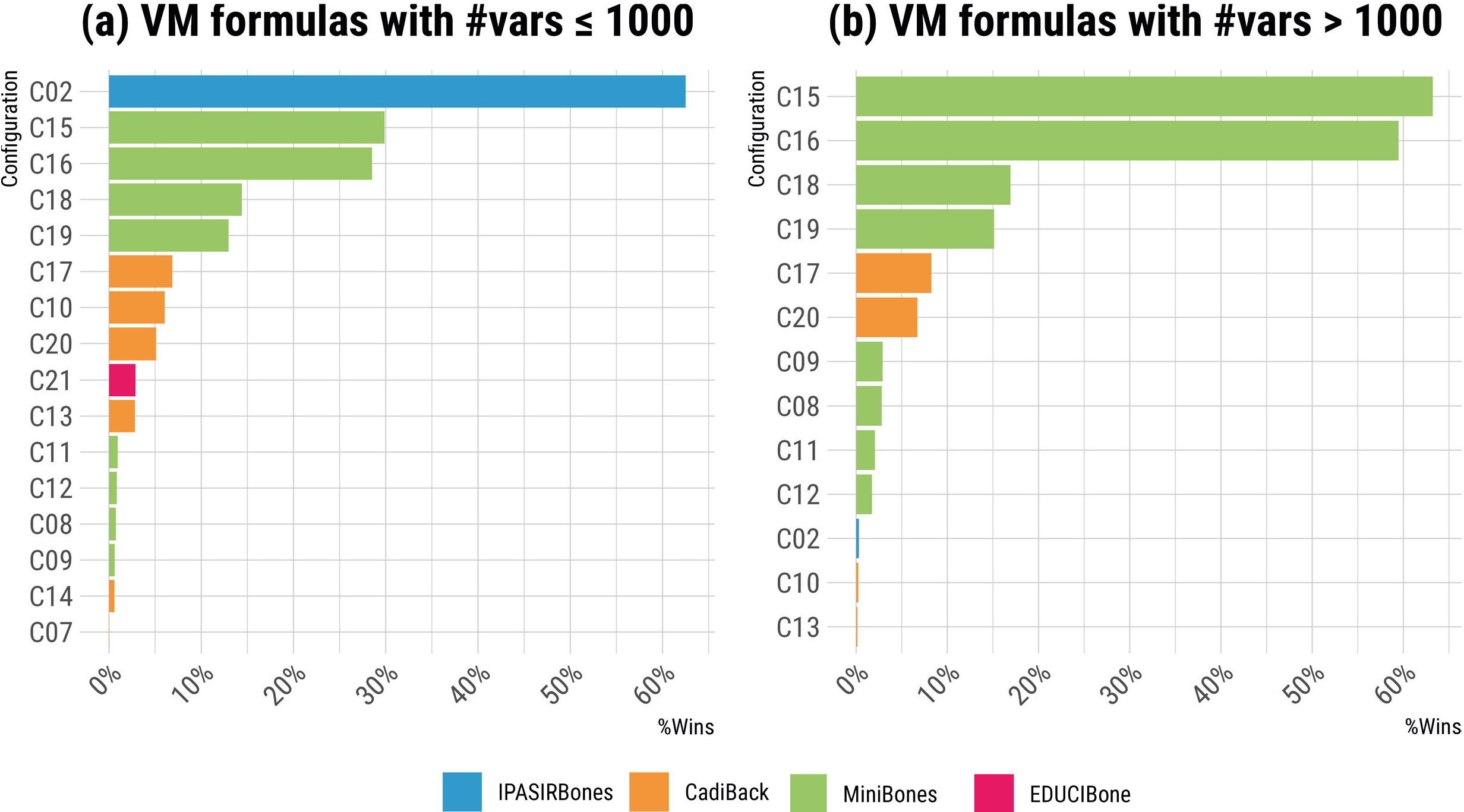}
    \caption{Ranking of the configurations that achieved the best times for the highest percentage of formulas in the dataset. (a) shows the top configurations for the formulas with 1,000 or fewer variables, and (b) for the formulas with more than 1,000 variables. Only configurations that won for at least one formula are depicted. For \tool{MiniBones}'s configurations, the figure considers the $k$'s that yielded the best time for each formula.}
    \label{fig:best-confs}
\end{figure*}

Fig. \ref{fig:best-confs} shows two histograms, (a) is focused on the VM formulas with 1,000 or fewer variables, while (b) is on the formulas with more than 1,000 variables.  The histograms list, on the $y$-axis, the configurations that achieved the best time for at least one formula; the $x$-axis shows, for those configurations, the percentage of formulas for which they got the shortest backbone computation time. Accordingly,
\begin{itemize}[leftmargin=*]
  \item Alg. \ref{alg:advanced-iterative}/\ref{alg:simplified-iterative} (C2) is the fastest for formulas with $\leq 1,000$ variables.
  \item Alg. \ref{alg:all-out} (C15 and C16) is the fastest for formulas with $> 1,000$ variables.
\end{itemize}

Remember from Table \ref{tab:config-summary} that most of \tool{MiniBones}’ configurations were tested on 21 values for $k$ ($k=5, 10, \ldots, 100,$ \#vars). Fig. \ref{fig:best-confs} represents the results obtained with the best $k$'s. Therefore, the green bars overestimate  \tool{MiniBones}’ performance. Nevertheless, all optimal configurations for large formulas were achieved with Alg.~\ref{alg:all-out}, including those utilizing \tool{CadiBack}'s automated $k$ selection (C17 and C20), which eliminates the overestimation mentioned before.

\begin{figure*}[htbp!]
    \centering
    \includegraphics[width=1\linewidth]{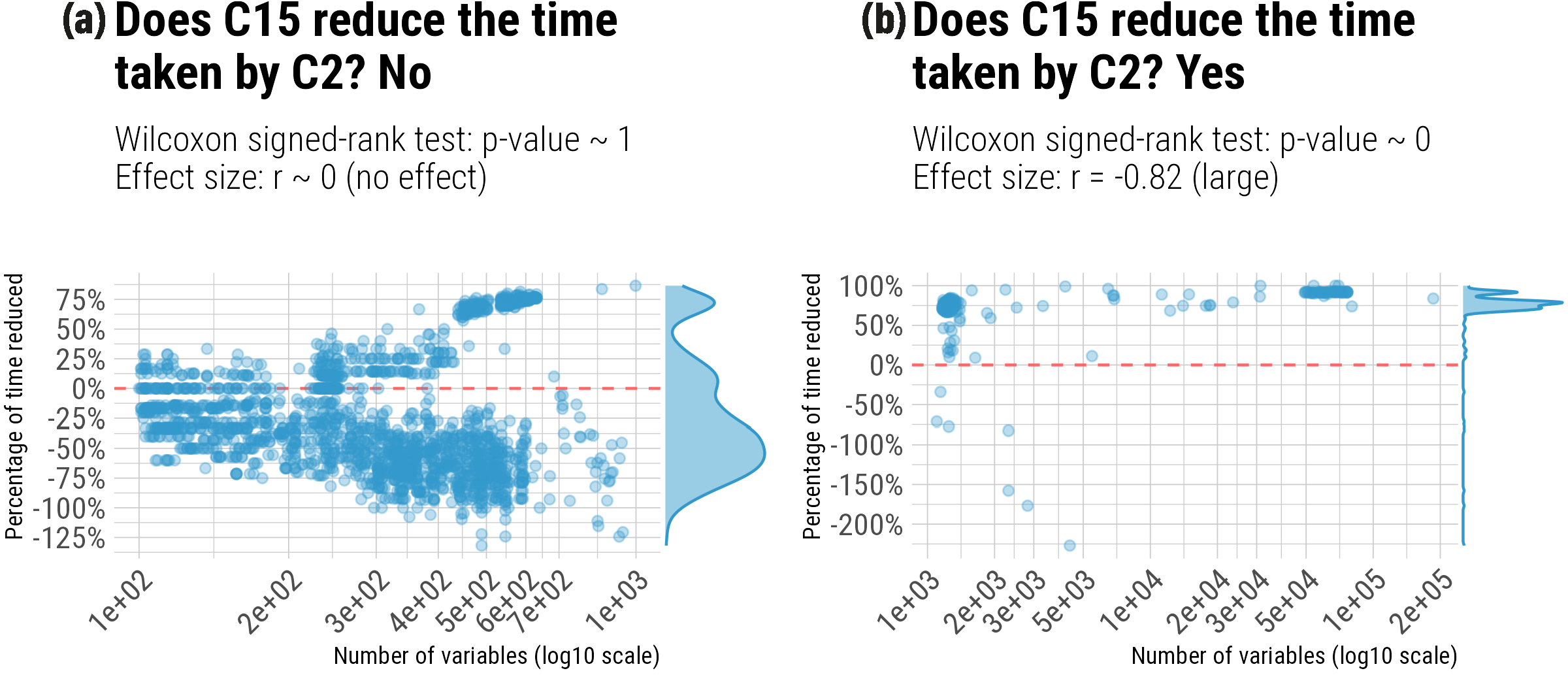}
    \caption{Percentage reduction in the runtime of C2 achieved by C15. Refer to the caption of Fig. \ref{fig:nclauses-vs-nvars} for a detailed explanation of the scatter and density plots.}
    \label{fig:C2-C15}
\end{figure*}

Fig. \ref{fig:C2-C15} compares in more detail the two winners, C2 and C15, using the same notation as in Fig. \ref{fig:c1_c2} (the comparisons between all pairs of configurations are available at our Zenodo repository). (a) proves that C15 worsens the results of C2 for formulas with $\leq 1,000$ variables, while (b) shows the contrary for formulas with $> 1,000$ variables.

Finally, it is important to note that the default configurations of \tool{MiniBones} and \tool{CadiBack} did not get the best results in the experiments:
\begin{itemize}[leftmargin=*]
    \item The default configuration for \tool{MiniBones} is C16 with $k$ = 100. However,  the best $k$ for C16 was 100 in only 14.68\% of the formulas.
    \item The default configuration for \tool{CadiBack} is C14, which was optimal in only 0.59\% of the formulas, all of which contained 1,000 or fewer variables.
\end{itemize}

\begin{center}
\begin{tcolorbox}[summary,title=\textbf{Answer to Q3:  \textit{Are there faster algorithms than Alg. 1 or Alg. 2/3?}}]
\textbf{Yes, there are.}
Alg. \ref{alg:all-out}, and to a much lesser extent Alg. \ref{alg:all-in}, demonstrate superior performance for formulas containing more than 1,000 variables (for formulas with $\leq$ 1,000 variables, Alg. \ref{alg:advanced-iterative}/\ref{alg:simplified-iterative} remains the fastest option). With optimal $k$ values, Alg. \ref{alg:all-out} can achieve reduction percentages exceeding 50\% for the majority of formulas, though such optimal values cannot be reliably estimated in practice. \tool{CadiBack}'s adaptive $k$ selection provides the best practical performance for large formulas. The default configurations of \tool{MiniBones} and \tool{CadiBack} do not perform well for VM formulas. 
\end{tcolorbox}
\end{center}

\subsubsection{Q4: Estimation of the Best $k$}

\begin{figure*}[htbp!]
    \centering
    \includegraphics[width=1\linewidth]{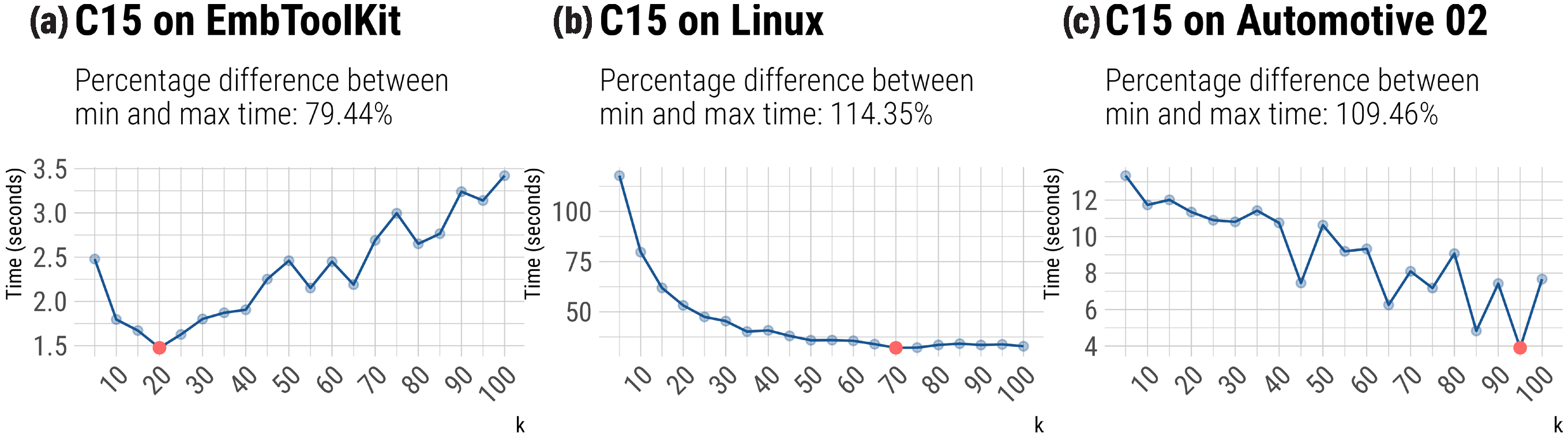}
    \caption{Time taken by C15 with different $k$'s for three formulas encoding the VMs of \tool{EmbToolkit}, \tool{Linux}, and \tool{Automotive 2.4}. Each data point indicates the seconds C15 needed ($y$-axis) for a specific chunk size $k$ ($x$-axis). The red point indicates the optimal $k$ that minimizes the time. }\label{fig:k_is_important}
\end{figure*}

Figure \ref{fig:k_is_important} illustrates an important problem of Algs. \ref{alg:all-in} and \ref{alg:all-out}: the parameter $k$ substantially influences algorithm performance, yet there is no systematic method for selecting the optimal $k$, as it varies depending on the specific formula. Each subfigure's subtitle indicates the percentage difference between the best $t_\mathrm{best}$ and the worst $t_\mathrm{worst}$ times, calculated as:
$$\frac{t_\mathrm{worst}-t_\mathrm{best}}{\big{ (}\frac{t_\mathrm{worst}+t_\mathrm{best}}{2}\big{)}}\cdot 100$$

\begin{figure*}[htbp!]
    \centering
    \includegraphics[width=1\linewidth]{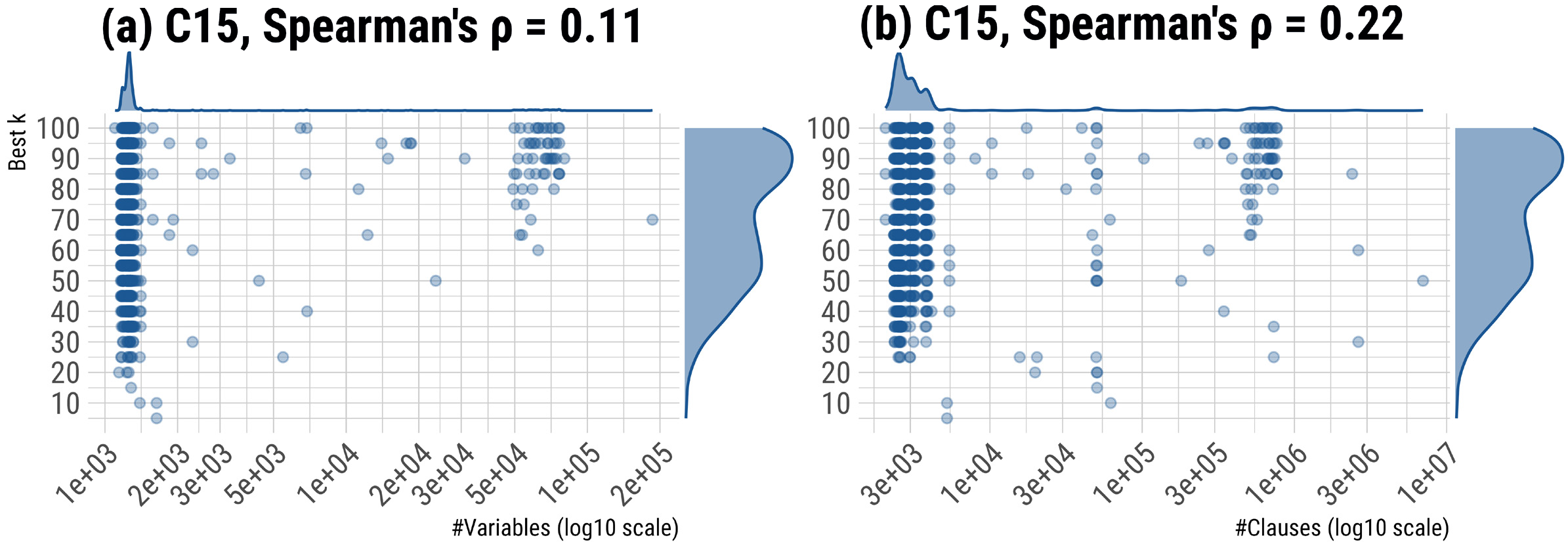}
    \caption{Spearman's $\rho$ correlation between the optimal $k$'s for C15 and the number of variables (a) and clauses (b) in the formulas (both correlations are identical when rounded to two decimal). Each point represents a formula, with the $x$-coordinate indicating its number of variables (a) or clauses (b), and the $y$-coordinate indicating its best $k$. Density diagrams at the top and right show the marginal distributions of the number of variables (a) and clauses (b), and the optimal $k$'s, respectively.}\label{fig:best_k_vs_num_vars_clauses}
\end{figure*}

The figure examines three formulas with more than 1,000 variables (\tool{EmbtoolKit} \cite{fernandez23} has 31,181 variables, \tool{Linux} \cite{fernandez23} has 186,059 variables, and \tool{Automotive 2.4} \cite{knuppel17} has 18,616 variables). The optimal $k$ varies widely across formulas. In \tool{EmbtoolKit}, C15 performance decreases as $k$ increases, with optimal performance at $k=20$. Conversely, in \tool{Linux} and \tool{Automotive 2.4}, C15 performance declines as $k$ decreases, with optimal values achieved at $k=70$ and $k=95$, respectively. These differences are not specific to C15 or the systems in Fig. \ref{fig:k_is_important}, but happened across all systems and configurations of Algs. \ref{alg:all-in} and \ref{alg:all-out} (refer to our Zenodo repository).

Fig. \ref{fig:best_k_vs_num_vars_clauses} provides more comprehensive detail, showing the correlation between the optimal $k$’s for each formula with $> 1,000$ variables and the number of variables (a) and clauses (b) in the formulas. The correlation was measured with Spearman's $\rho$, which accounts for both linear and non-linear correlations ($|\rho| < 0.2$ is typically considered negligible \cite{spearman-corr}). The marginal distributions on the right further reveal that the optimal $k$ values are distributed broadly. This distribution suggests that the optimal $k$ is highly variable and specific to each formula. The analogous figures for the remaining configurations of Algs. \ref{alg:all-in} and \ref{alg:all-out} with fixed $k$ (C8, C9, C11, C12, C16, C18, and C19) are available in our Zenodo repository and demonstrate that weak correlations occur in all configurations. As a result, neither the number of variables nor the number of clauses are effective predictors for the optimal $k$.

\begin{figure}[htbp!]
    \centering
    \includegraphics[width=1\linewidth]{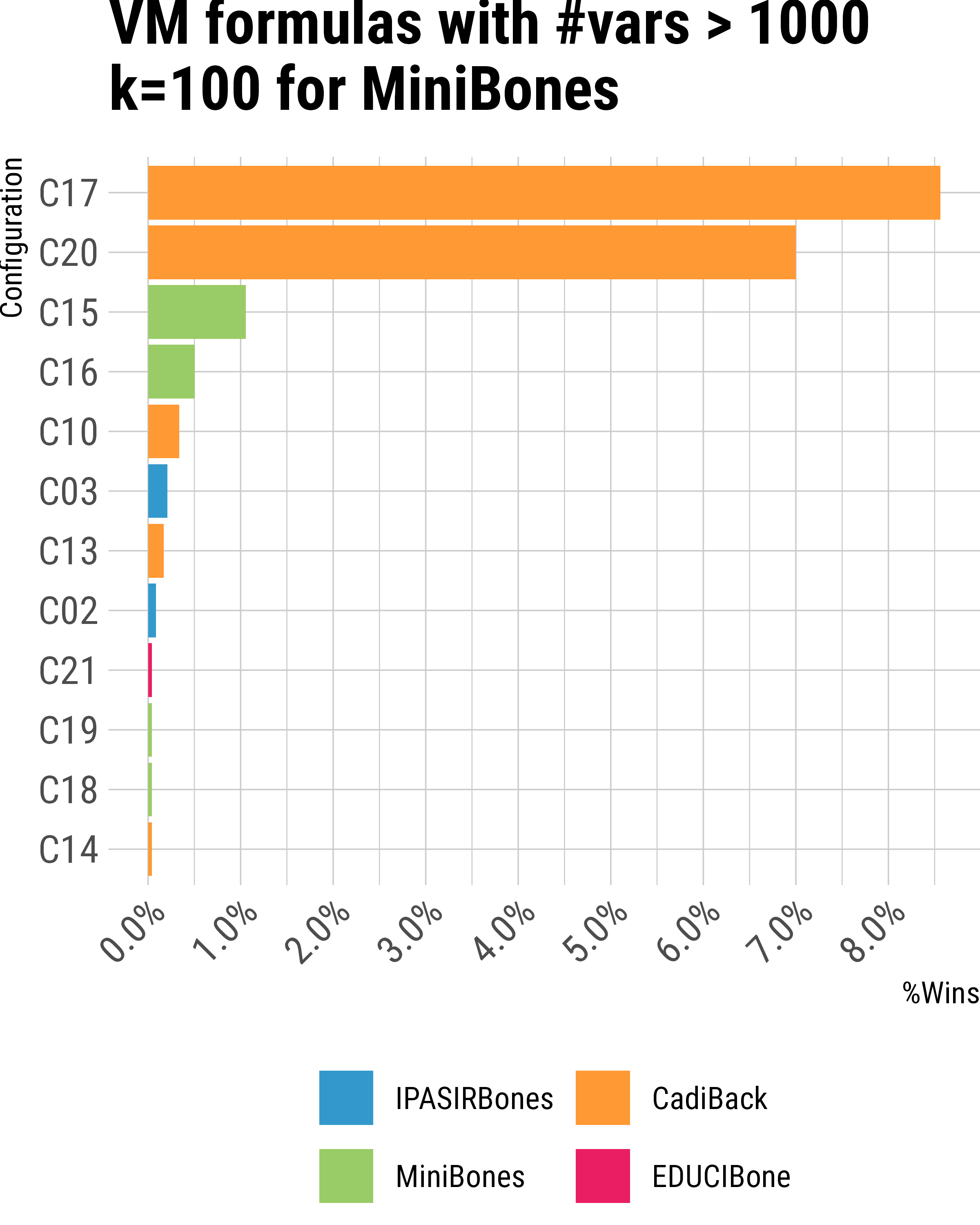}
    \caption{Ranking of the configurations that achieved the best times for the highest percentage of formulas with $>1,000$ variables. \tool{MiniBones}’ $k$ was set to 100 (its default value for C16).}\label{fig:best_confs_default_values}
\end{figure}

Fig. \ref{fig:best_confs_default_values} ranks the best configurations for formulas containing more than 1,000 variables, utilizing the default $k=100$ for \tool{MiniBones}. The results differ significantly from those in Figure \ref{fig:best-confs}.(b), which employed the optimal values of $k$. As there is no reliable method for estimating a superior $k$ than the default, this Fig. \ref{fig:best_confs_default_values} reflects more pragmatically the algorithms' performance. 

Table \ref{tab:times-hardest} presents a comparison of the execution times between the most effective configuration (C17 with adaptive $k$) and the best idealistic configuration (C15 with optimal $k$ values) for the most challenging formulas in the benchmark. Although C17 occasionally outperforms C15, this outcome is uncommon.

\begin{figure*}[htbp!]
    \centering
    \includegraphics[width=1\linewidth]{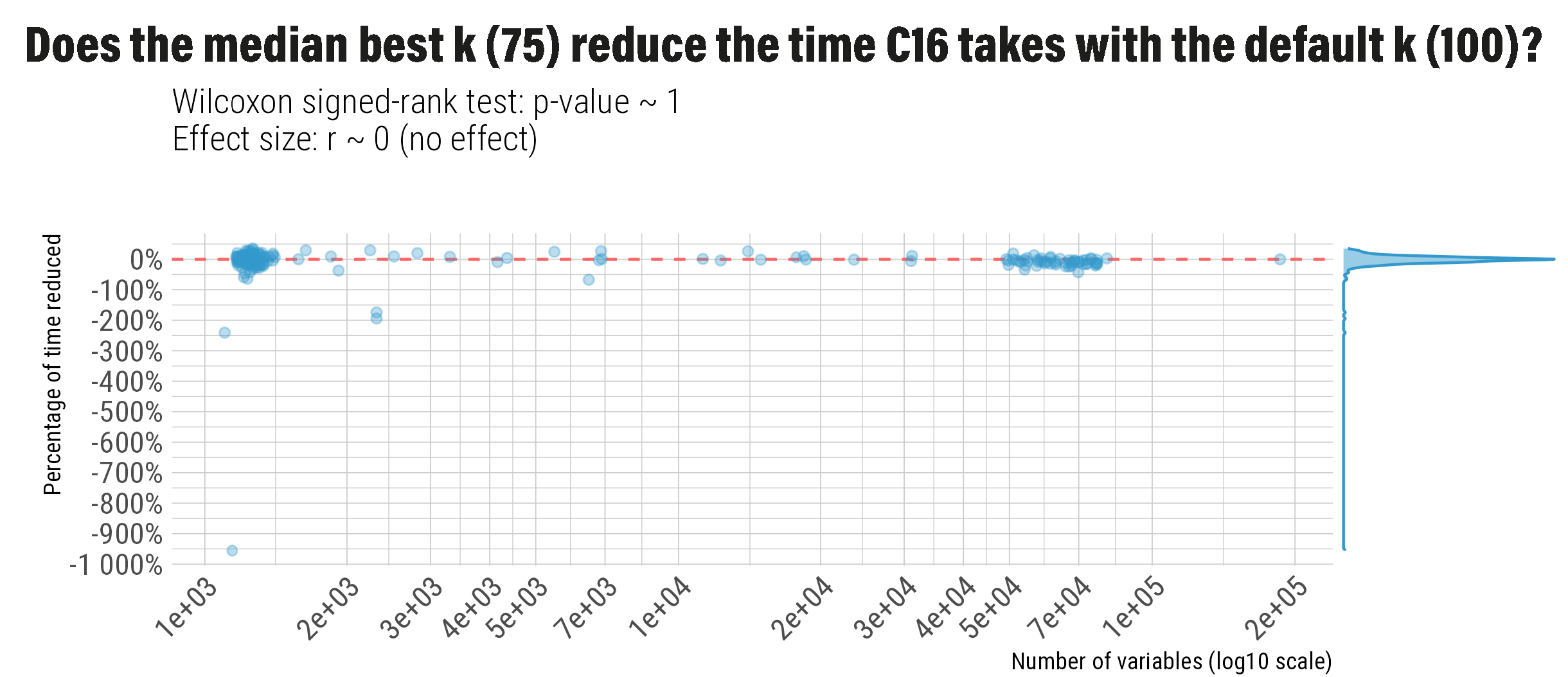}
    \caption{The median best $k$ cannot reduce the time C16 takes with the default $k$.}\label{fig:median_over_default}
\end{figure*}

\newpage
We sought to enhance \tool{MiniBones} performance by replacing its default $k=100$ with a centrality measure, specifically the median of the best $k$ values (the median was selected over the mean because the distribution of the best $k$ values is asymmetric; see the right density diagrams in Fig. \ref{fig:best_k_vs_num_vars_clauses}). Table \ref{tab:best_medians} summarizes the median best $k$'s for \tool{MiniBones}’ configurations. However, Fig. \ref{fig:median_over_default} shows that this does not work (because the best $k$’s have too much variance).

\begin{table}[htbp!]
\caption{Comparison between the time taken by C17 (adaptive $k$) and C15 (with optimal $k$’s) for the hardest formulas in the benchmark.}\label{tab:times-hardest}
\begin{footnotesize}
\centering
\begin{tabular}{|l|r|r|r|r|}
  \hline
  VM & \#vars. & \#clauses & C17 & C15 \\
  & & & secs. & opt. $k$ \\
  & & &  & secs. \\
   \hline \hline
  \tool{Linux} \cite{fernandez23} & 186,059 & 527,240 & 23.06 & 32.10\\ \hline
  \tool{Linux} \cite{Pett19} & 76,487 & 767,040 & 62.42 & 26.27 \\ \hline
  \tool{Coreboot} \cite{fernandez23} & 69,837 & 345,749 & 186.78 & 16.24 \\ \hline
  \tool{Freetz} \cite{fernandez23} & 67,546 & 240,767 & 40.57 & 8.08 \\ \hline
  \tool{Embtoolkit} \cite{fernandez23} & 31,181 & 78,545 & 2.37 & 1.48 \\ \hline
  \tool{Automotive} & 18,616 & 350,120 & 4.52 & 3.90\\
  2.4 \cite{knuppel17} & & & & \\ \hline
  \tool{Buildroot} \cite{Berger13} & 14,910 & 45,603 & 0.11 & 0.27 \\ \hline
  \tool{uCLinux} \cite{Berger13} & 11,254 & 31,637 & 0.06 & 0.16 \\ \hline
\end{tabular}
\end{footnotesize}
\end{table}

\begin{table}[htbp!]
\caption{Median of the best $k$’s for \tool{MiniBones}’s configurations.}\label{tab:best_medians}
\centering
\begin{tabular}{|l|r|}
  \hline
\textbf{Config.}  & \textbf{Median of}\\
\textbf{Id} & \textbf{the best $k$'s} \\
  \hline
  \hline
    C8 &   5 \\  \hline
    C9 &   5 \\ \hline
    C11 &   5 \\ \hline
    C12 &   5 \\ \hline
    C15 &  80 \\ \hline
    C16 &  75 \\ \hline
    C18 &  85 \\ \hline
    C19 &  85 \\ \hline
\end{tabular}
\end{table}

\begin{figure*}[htbp!]
    \centering
    \includegraphics[width=1\linewidth]{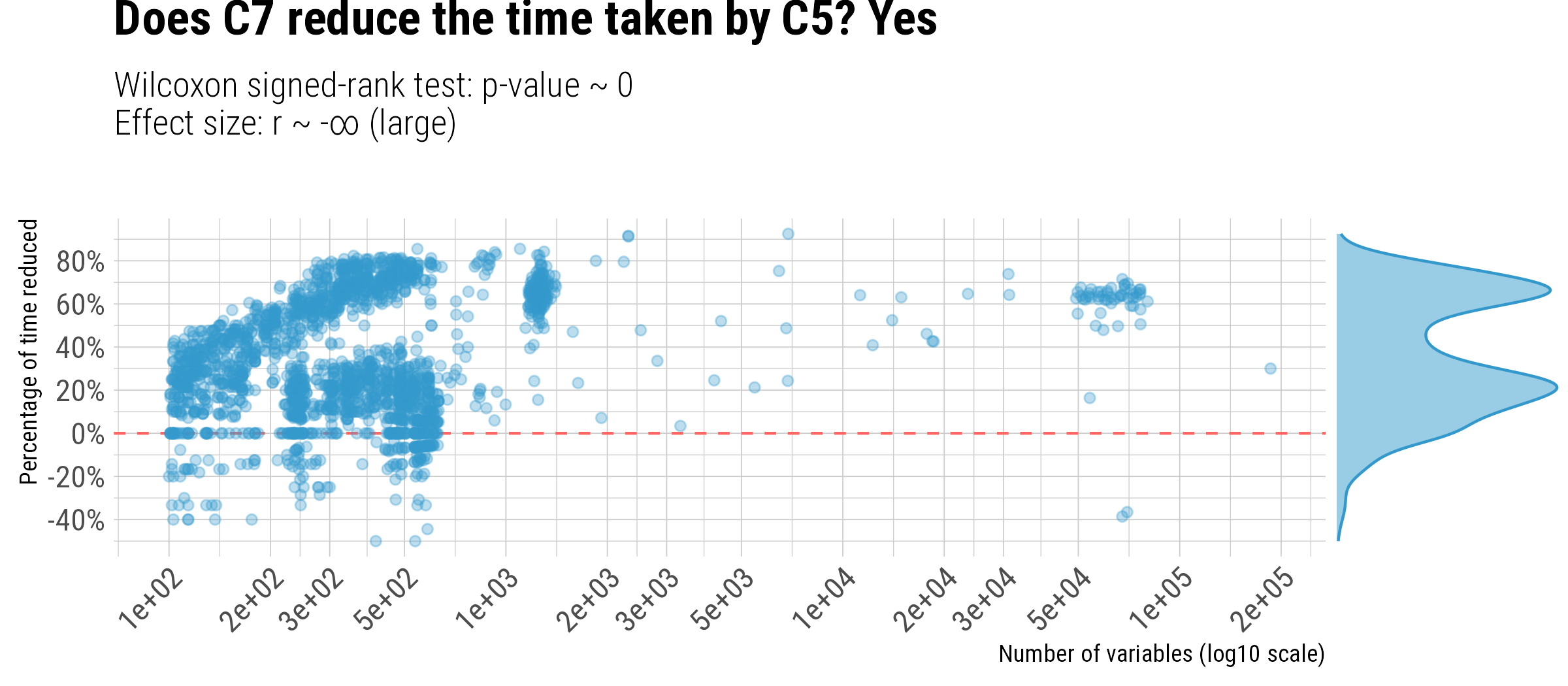}
    \caption{TIP improves Alg. \ref{alg:advanced-iterative}/\ref{alg:simplified-iterative}'s performance when it is combined with filtering RLs. In the remaining cases, it has no influence.}\label{fig:tip-influence}
\end{figure*}

\begin{center}
\begin{tcolorbox}[summary,title=\textbf{Answer to Q4: \textit{Is it possible to infer the optimal $k$ for Algs. 4 and 5 from easily computable formula features, e.g., the number of variables or clauses?}}]
\textbf{No, it is not.}
The experiments show that $k$ is specific to each formula and varies randomly with the number of variables or clauses, so these cannot work as reliable estimators for $k$.
\end{tcolorbox}
\end{center}

\subsubsection{Q5: Influence of UC injection}

The impact of injecting backbone literals after their detection as unit clauses in the formula was assessed by comparing the following configuration pairs: (i) C2 \textit{vs.} C3 and C5 \textit{vs.} C6 for Alg. \ref{alg:advanced-iterative}/\ref{alg:simplified-iterative}, (ii) C8 \textit{vs.} C9 and C11 \textit{vs.} C12 for Alg. \ref{alg:all-in}, and (iii) C15 \textit{vs.} C16 and C18 \textit{vs.} C19 for Alg. \ref{alg:all-out}. None of these comparisons revealed any influence (neither positive nor negative).



\begin{center}
\begin{tcolorbox}[summary,title=\textbf{Answer to Q5: \textit{Does UC injection influence the algorithms’ performance?}}]
\textbf{No, it does not.}
\end{tcolorbox}
\end{center}

\subsubsection{Q6: Influence of TIP}

As mentioned in Section \ref{sec:filtering}, the \tool{CaDiCal} SAT solver implements TIP, an improved and transparent approach to injecting backbone literals in the formula as they are discovered. The impact of TIP was assessed by comparing the following configuration pairs: (i) C2 \textit{vs.} C4 and C5 \textit{vs.} C7 for Alg. \ref{alg:advanced-iterative}/\ref{alg:simplified-iterative}, (ii) C8 \textit{vs.} C10 and C11 \textit{vs.} C13 for Alg. \ref{alg:all-in}, and (iii) C15 \textit{vs.} C17 and C18 \textit{vs.} C20 for Alg. \ref{alg:all-out}. All comparisons showed no influence, except for the one shown in Fig. \ref{fig:tip-influence}: TIP improves Alg. \ref{alg:advanced-iterative}/\ref{alg:simplified-iterative}'s performance when it is combined with filtering RLs.

\begin{center}
\begin{tcolorbox}[summary,title=\textbf{Answer to Q6: \textit{Does TIP influence the algorithms’ performance?}}]
\textbf{Yes, it does}, but only for Alg. \ref{alg:advanced-iterative}/\ref{alg:simplified-iterative} when it is combined with filtering RLs.
\end{tcolorbox}
\end{center}

\begin{figure*}[htbp!]
    \centering
    \includegraphics[width=1\linewidth]{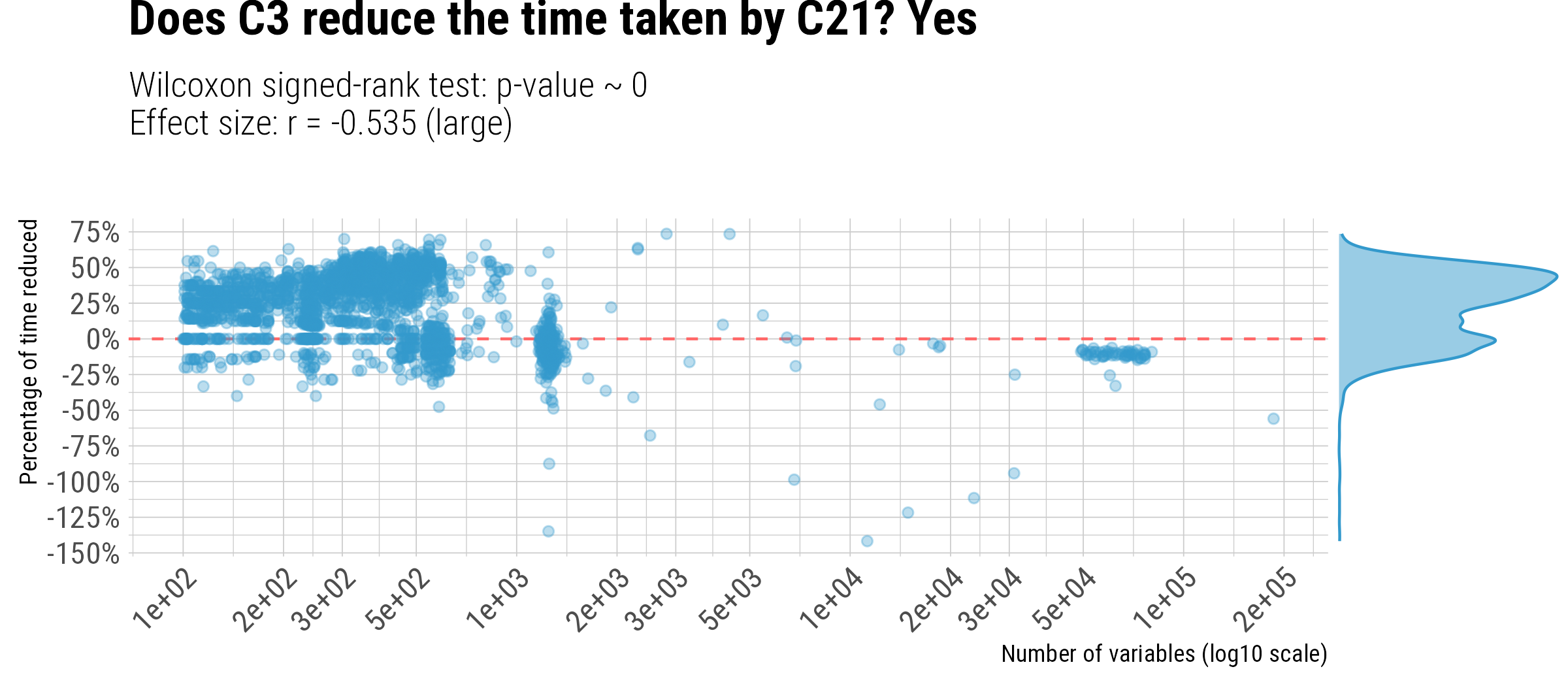}
    \caption{COV, WHIT, and 2LEN negatively impacts the performance of Alg. \ref{alg:advanced-iterative}/\ref{alg:simplified-iterative}}\label{fig:educibone-influence}
\end{figure*}

\subsubsection{Q7: Influence of RLs}

The impact of literal filtering by using RLs was assessed by comparing the following configuration pairs: (i) C2 \textit{vs.} C5, C3 \textit{vs.} C6, and C4 \textit{vs.} C7 for Alg. \ref{alg:advanced-iterative}/\ref{alg:simplified-iterative}, (ii) C8 \textit{vs.} C11, C9 \textit{vs.} C12, and C10 \textit{vs.} C13 for Alg. \ref{alg:all-in}, and (iii) C15 \textit{vs.} C18, C16 \textit{vs.} C19, and C17 \textit{vs.} C20 for Alg. \ref{alg:all-out}. None of these comparisons revealed any influence (neither positive nor negative).

\begin{center}
\begin{tcolorbox}[summary,title=\textbf{Answer to Q7: \textit{Does literal filtering with RLs influence the algorithms’ performance?}}]
\textbf{No, it does not.}
\end{tcolorbox}
\end{center}

\subsubsection{Q8: Influence of COV, WHIT and 2LEN}

The combined application of COV, WHIT, and 2LEN was assessed by comparing C21 \textit{vs.} C3. Fig.~\ref{fig:educibone-influence} shows that these heuristics have a statistically significant and substantial negative effect on Alg. \ref{alg:advanced-iterative}/\ref{alg:simplified-iterative}. 

Unlike Figs. \ref{fig:c1_c2}, \ref{fig:C2-C15}, and \ref{fig:tip-influence}, Fig.~\ref{fig:educibone-influence} investigates whether C3 reduces the time required by C21, rather than the reverse. The rationale for this analysis is as follows. For each pair of configurations $C$ and $C'$, our Zenodo repository includes two figures: (i) $C$ \textit{vs.} $C'$ and (ii) $C'$ \textit{vs.} $C$. The first figure ($C$ \textit{vs.} $C'$) depicts how much $C'$ reduces the execution time of $C$ and tests the null hypothesis $H_0$: ``$C'$ runtime is equal to or longer than $C$ runtime'', against the alternative hypothesis $H_a$: ``$C'$ runtime is shorter than $C$ runtime''. The second figure ($C'$ \textit{vs.} $C$) represents the reverse comparison.

In inferential testing, the rejection of $H_0$ based on experimental data leads to the acceptance of $H_a$. However, the converse is not true~\cite{Field12}: if the data do not support rejecting $H_0$, it cannot be inferred that $H_0$ is \code{true} and $H_a$ \code{false}. In other words, the test is inconclusive. 

Although the figure for C3 \textit{vs.} C21 in Zenodo suggests that C21 worsens C3, the non-rejection of $H_0$ does not support any definitive inferential conclusion. Consequently, the reverse comparison (C21 versus C3) is presented in Fig.~\ref{fig:educibone-influence}, which does demonstrate that C21 significantly worsens C3 ($p$-value close to 0 and effect size $|r| > 0.5$).

\begin{center}
\begin{tcolorbox}[summary,title=\textbf{Answer to Q8: \textit{Does the combined literal filtering with COV, WHIT, and 2LEN influence the algorithms’ performance?}}]
\textbf{Yes, it does}, but quite negatively.
\end{tcolorbox}
\end{center}

\subsection{Discussion}

\subsubsection{Key Findings}

Our experiments reveal that formulas derived from VMs have fundamentally different structural properties compared to those used in SAT competitions. For formulas with 1,000 or fewer variables, Alg.~\ref{alg:advanced-iterative}/\ref{alg:simplified-iterative} consistently outperforms chunking algorithms. For larger formulas, Alg.~\ref{alg:all-out} with \tool{CadiBack}'s adaptive $k$ selection provides the best practical performance. Additionally, most heuristics either have no effect (injecting backbone literals in the formula and filtering RLs) or negatively impact performance on VM formulas (COV, WHIT, and 2LEN).

A key finding concerns the chunk size parameter $k$ in Alg.~\ref{alg:all-out}. With ``oracle knowledge'' of optimal $k$ values, \tool{MiniBones} would achieve the best performance for formulas with more than 1,000 variables. However, this theoretical advantage cannot be exploited in practice. The distribution of optimal $k$ values exhibits three characteristics that prevent reliable estimation: (i)~weak correlation with  formula features that are cheap to compute, such as the number of variables or clauses (Spearman's $\rho < 0.2$), (ii)~high variance across formulas, and (iii)~strong asymmetry that renders centrality measures like the median ineffective (Fig.~\ref{fig:best_k_vs_num_vars_clauses}). Consequently, with its default $k=100$, \tool{MiniBones} is outperformed by \tool{CadiBack}'s configurations using adaptive $k$ selection (C17 and C20), as shown in Fig.~\ref{fig:best_confs_default_values}.

Taken together, these results indicate that the available algorithms are fast enough for practical VM analysis but could be improved with better mechanisms for tuning $k$. Nevertheless, even for the most challenging model in our benchmark, the \tool{Linux Kernel} from~\cite{fernandez23} with 186,059 variables and 527,240 clauses, C17  computed the backbone in just 23.06 seconds (see Table \ref{tab:times-hardest}).

\subsubsection{Impact}

\textbf{For tool developers.} Our results suggest several improvements to backbone computation tools:

\begin{itemize}[leftmargin=*]
    \item \textit{Implement size-based algorithm selection.} Tools should automatically select Alg.~\ref{alg:advanced-iterative}/\ref{alg:simplified-iterative} for formulas with $\leq$1,000 variables and Alg.~\ref{alg:all-out} for larger formulas. This hybrid approach would outperform any single algorithm across the full range of VM sizes.

    \item \textit{Reconfigure default settings.} The default configurations of \tool{MiniBones} ($k$=100 with Alg.~\ref{alg:all-out}) and \tool{CadiBack} (Alg.~\ref{alg:all-in} with $k$=$\infty$) are suboptimal for VM formulas. To improve performance on large VMs, we recommend configuring \tool{CadiBack} to use Alg.~\ref{alg:all-out} with automatic $k$ selection (C17), which, as described in Section \ref{sec:tools},  starts with $k=1$ and multiplies $k$ by 10 in each loop iteration, provided the all-in/all-out condition is satisfied. If this condition is not met, $k$ is reset to 1.

    \item \textit{Disable literal filtering for VM analysis.} The filtering heuristics COV, WHIT, and 2LEN harm performance (Fig.~\ref{fig:educibone-influence}), while filtering RLs have no measurable effect. Disabling these mechanisms simplifies implementation and improves runtime.
\end{itemize}

\textbf{For practitioners.} Based on our findings, we recommend:

\begin{itemize}[leftmargin=*]
    \item For small to medium VMs ($\leq$1,000 variables), use tools implementing Alg.~\ref{alg:advanced-iterative}/\ref{alg:simplified-iterative}, such as \tool{IPASIRBones}.

    \item For large VMs ($>$1,000 variables), use \tool{CadiBack}~\cite{Biere2023_CadiBack} with Alg.~\ref{alg:all-out} and automatic $k$ selection. Although \tool{MiniBones} with optimal $k$ values can achieve greater runtime reductions, determining these optimal values is not feasible in practice at the moment. \tool{CadiBack}'s adaptive $k$ mechanism provides robust performance without requiring manual parameter tuning.
\end{itemize}

\textbf{For researchers.} Our study highlights the importance of benchmarking on realistic VM formulas rather than relying exclusively on SAT competition benchmarks. Algorithm performance can differ dramatically between these domains, and results from one may not generalize to the other. Our findings suggest two directions for future research:

\begin{itemize}[leftmargin=*]
    \item \textit{Develop methods for estimating optimal $k$ values.} Future work might explore alternative predictive features (e.g., formula-structure metrics exploiting the distinctive clause patterns of VM formulas) or machine-learning approaches.

    \item \textit{Improve adaptive $k$ selection mechanisms.} \tool{CadiBack}'s automatic $k$ selection outperforms fixed $k$ values in practical scenarios (Fig.~\ref{fig:best_confs_default_values}), yet Fig.~\ref{fig:best-confs}(b) shows substantial room for improvement. Future work should analyze whether more sophisticated adaptive strategies could achieve performance closer to the optimal $k$’s.
\end{itemize}

\subsubsection{Limitations}

Three \textit{types of validations} were considered in our empirical evaluation:
\begin{enumerate}[(a), wide, labelwidth=!, labelindent=0pt]
    \item \textbf{Construct validity} refers to the extent to which the intended constructs are accurately measured. The benchmark formulas are derived from established sources~\cite{Sundermann24benchmark,fernandez23} and represent real configurable systems across multiple domains, including operating systems (\tool{Linux}, \tool{BusyBox}, \tool{uClibc}), embedded systems (\tool{Buildroot}, \tool{EmbToolkit}, \tool{Freetz}), firmware (\tool{Coreboot}), and automotive software. The CNF translations were generated using two different tools (\tool{TraVarT}~\cite{Feichtinger21} and \tool{KconfigSampler}~\cite{fernandez23}). However, although the benchmark encompasses diverse systems and translations, it may over-represent \tool{Kconfig}-based systems and \tool{UVL} models \cite{benavides2024uvl}, and may not capture all structural characteristics of variability models in other specification languages.

    \item \textbf{Internal validity} concerns whether the observed performance differences can be attributed to the algorithms rather than confounding factors. Several measures ensured fair comparisons. First, to avoid bias from programming language differences, we implemented Algs.~\ref{alg:naive-iterative} and \ref{alg:advanced-iterative}/\ref{alg:simplified-iterative} in C++ (\tool{IPASIRBones}), matching the implementation language of \tool{MiniBones}, \tool{CadiBack}, and \tool{EDUCIBone}. Second, all experiments were run on the same hardware (an HP Proliant server with an Xeon E5-2660v4) to ensure consistent conditions. Third, each configuration was run three times per formula, and the median time was used for analysis. This approach provided robustness against outliers from caching effects or background processes. Fourth, the $k$ parameter was tested comprehensively (ranging from 5 to 100, including $k$=$\#$vars and automatic selection) to avoid bias from arbitrary parameter choices. Fifth, statistical comparisons used paired Wilcoxon signed-rank tests, comparing algorithms on the same formulas, with effect sizes reported following Cohen's guidelines~\cite{Cohen88}. Despite these precautions, our results remain tied to specific SAT solver implementations (\tool{MiniSat} 2.2 and \tool{CaDiCaL}). 

    \item \textbf{External validity} concerns whether results generalize beyond this study. Our benchmark of 2,371 formulas from diverse real-world systems provides broad coverage, and our findings were consistent across this entire dataset. However, generalization has limits: (i) our results explicitly demonstrate that findings do not transfer between domains, i.e., algorithms that excel on SAT competition formulas perform differently on VM formulas due to their distinct structural properties; (ii) formulas from other domains may exhibit different clause structures where our conclusions would not apply; (iii) SAT solver technology evolves rapidly, and future algorithmic advances could alter relative performance rankings, though the structural characteristics of VM formulas (approximately 98\% of clauses with $\leq$2 literals) are unlikely to change; and (iv) while we tested formulas ranging up to 186,059 variables (\tool{Linux Kernel}), extremely large future VMs could potentially behave differently.
\end{enumerate} 

\section{Conclusions}\label{sec:conclusions}

This work has presented the first empirical evaluation of backbone algorithms that incorporates the latest advances from the SAT community and focuses on formulas derived from real-world VMs. As a result, we have provided clear and actionable guidelines. For formulas with 1,000 or fewer variables, Alg.~\ref{alg:advanced-iterative}/\ref{alg:simplified-iterative} (iterative with solution filtering) consistently outperforms alternative approaches. For larger formulas, Alg.~\ref{alg:all-out} (chunked core-based) with an optimal chunk size $k$ can reduce runtime by more than 50 percent. However, identifying optimal $k$ values is not feasible in practice. Consequently, \tool{CadiBack}'s adaptive $k$ selection provides the best practical performance. Filtering heuristics either have no measurable effect or negatively impact performance on variability model formulas. Therefore, tool developers are advised to implement size-based algorithm selection and disable literal filtering. These findings have significant practical implications, as backbone computation has proved tractable even for the largest real-world systems. For instance, one \tool{CadiBack} configuration computed the \tool{Linux Kernel}'s backbone (with 186,059 variables) in just 23.06 seconds.

Our study has also pointed out new research areas. Future work should explore methods to predict good chunk sizes $k$ for Alg. \ref{alg:all-out}, such as by analyzing the formula's structure, employing machine learning, or developing improved adaptive methods.

In summary, this unified overview of backbone algorithms has introduced the SPL community to advanced SAT techniques that had not yet been explored in SPL engineering. This work is expected to promote wider use of these algorithms in VM analysis tools.



\vspace{0.5cm}
\noindent \textbf{Funding.} This work is funded by FEDER/Spanish Ministry of Science, Innovation and Universities (MCIN)/Agencia Estatal de Investigacion (AEI) under grant COSY (PID2022-142043NB-I00), IRIS (PID2021-122812OB-I00), SAVIA (PID2024-159945NB-I00), and Data-PL (PID2022-138486OB-I00); and by Junta de Andalucía under grant DUNE (DGP\_PIDI\_ 2024\_00092). 

\vspace{0.5cm}
\noindent \textbf{Data availability.} The data used in this research is publicly available at: \url{https://doi.org/10.5281/zenodo.18494650}

\vspace{0.5cm}
\noindent \textbf{Declarations.} The authors have no competing interests to declare that are relevant to the content of this article.


\end{document}